\documentclass[aps,prx,reprint,superscriptaddress,nofootinbib,longbibliography,floatfix]{revtex4-2}
\usepackage[T1]{fontenc}
\usepackage[utf8]{inputenc}
\usepackage{amsmath,amssymb,bm}
\usepackage{graphicx,booktabs,xcolor}
\usepackage[colorlinks=true,linkcolor=blue,citecolor=blue,urlcolor=blue]{hyperref}

\newcommand{\ket}[1]{\lvert #1\rangle}
\newcommand{\bra}[1]{\langle #1\rvert}
\newcommand{\braket}[2]{\langle #1\vert #2\rangle}
\newcommand{\Tr}{\operatorname{Tr}}
\newcommand{\Rea}{\operatorname{Re}}
\newcommand{\Ima}{\operatorname{Im}}
\newcommand{\id}{\mathbb{I}} 
\newcommand{\thv}{\boldsymbol{\theta}}
\newcommand{\vv}{\boldsymbol{v}}

\usepackage{microtype} 

\begin{document}
\title{Thermal state preparation with shallow quantum circuits}
\author{Cenk T\"uys\"uz}
\email{cenk.tuysuz@cern.ch}
\affiliation{European Organisation for Nuclear Research (CERN), Geneva, Switzerland}
\author{Simon Deconihout}
\affiliation{\'{E}cole Polytechnique F\'{e}d\'{e}rale de Lausanne (EPFL), Lausanne, Switzerland}
\author{Enrico Rinaldi}
\affiliation{Quantinuum, Partnership House, Carlisle Place, London SW1P 1BX, United Kingdom}
\affiliation{RIKEN Center for Quantum Computing (RQC), RIKEN, Wako, Saitama 351-0198, Japan}
\affiliation{School of Mathematical Sciences, Queen Mary University of London, Mile End Road, London, E1 4NS, UK}
\author{Michele Grossi}
\affiliation{European Organisation for Nuclear Research (CERN), Geneva, Switzerland}

\date{\today}

\begin{abstract}
Preparing thermal states on noisy quantum hardware requires circuits that remain shallow at low temperatures. Even when accurate classical state representations are available, compiling them into suitable circuits can be costly. We introduce PP-VarQITE, combining Pauli propagation with variational quantum imaginary-time evolution to determine circuit parameters entirely classically, with target temperatures encoded in the angles of a fixed circuit architecture. In numerical studies of Ising and frustrated Heisenberg models, we recover low-temperature energies and correlations using up to 100 qubits. Across the frustrated chain and two-dimensional lattice, our circuits are approximately two to ten times shallower than the lowest-depth circuits from sequential tensor-network compilation. These shallow circuits enable us to prepare a 48-site frustrated chain and a $7\times7$ Ising lattice using 96 and 98 qubits on the Quantinuum Helios trapped-ion quantum processor. We reproduce the development of dimer order and lattice-wide correlations from high to low temperatures, with the lattice circuits requiring at most 42 two-qubit layers. Through classical circuit design, we demonstrate low-temperature state preparation at circuit sizes beyond the practical reach of statevector simulation.
\end{abstract}

\maketitle

\section{Introduction}\label{sec:introduction}

\begin{figure*}[t]
 \centering
  \includegraphics[width=\linewidth]{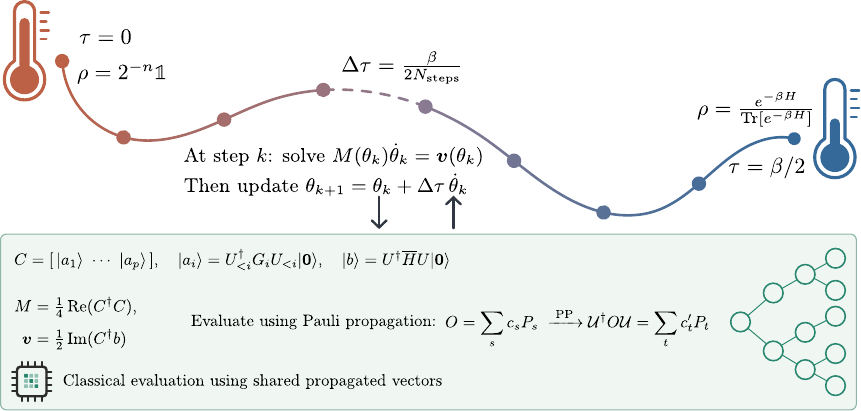}
 \caption{\textbf{Summary of the PP-VarQITE framework.}
Starting from an infinite-temperature purification, we determine the circuit angles by solving $M\dot{\thv}=\vv$ and applying Euler updates with $\Delta\tau=\beta/(2N_{\mathrm{steps}})$. The green box shows how Pauli propagation supplies the shared vectors used to assemble the Gram metric $M$ and evolution vector $\vv$ on a classical computer.}
\label{fig:workflow}
\end{figure*}

A central challenge in quantum simulation is to prepare correlated many-body states at controlled temperatures within the circuit depth available on noisy hardware. Evolving these states in real time gives access to finite-temperature dynamical correlation functions and excitation spectra, which characterize the response of interacting systems to external perturbations~\cite{Sun2021,Lee2022}. On noisy hardware, state preparation and real-time evolution share a limited circuit budget, so the depth required to initialize the system constrains the subsequent dynamics. When an approximate thermal state is sufficiently accurate for the observables of interest, reducing the depth of its state preparation circuit can allow longer real-time evolution before accumulated errors obscure the dynamical signal.

Low-temperature states are also useful for ground-state algorithms because the ground-state population increases as the temperature decreases. In sample-based Krylov diagonalization, measurements of real-time-evolved reference states determine a subspace in which the Hamiltonian is diagonalized classically~\cite{Yu2025}. For a pure reference state, larger ground-state overlap improves the convergence and sampling bounds under the method's spectral and sparsity assumptions, and can therefore reduce the resources required to estimate the ground-state energy.

The difficulty of obtaining these states depends on temperature as well as on the Hamiltonian. At sufficiently high temperatures, local Hamiltonians with bounded interaction strengths and connectivity admit polynomial-time classical algorithms for sampling Gibbs-state measurement outcomes and constructing mixtures of product states~\cite{Yin2023,Bakshi2024}. At lower temperatures, quantum Monte Carlo methods compute equilibrium observables for unfrustrated models~\cite{Sandvik1991}, but the sign problem obstructs frustrated and fermionic systems~\cite{Troyer2005}. Matrix product states (MPS) and thermal tensor-network methods extend classical calculations into the low-temperature regime~\cite{Feiguin2005,Chen2018,Li2023}, with their cost governed by the bond dimensions needed to capture the correlations. Thermal-state simulations based on Pauli expansions are also effective at high temperatures, but the retained operator representations and the associated classical cost grow as the temperature decreases~\cite{GomezLurbe2026,Rudolph2026}.

Quantum algorithms provide correctness and convergence guarantees for Gibbs state preparation in specified Hamiltonian and temperature regimes~\cite{Rouze2026,Ding2026}, but realizing these guarantees can require circuit depths beyond those available on noisy processors. For noisy devices, dissipative protocols approximate thermalization through weak system-bath coupling~\cite{Lloyd2026,Ong2026}, while adiabatic schedules evolve an initial product thermal state under an interpolating Hamiltonian~\cite{Granet2026}. Variational and quantum tensor-network methods instead approximate thermal states within shallower parametrized circuits~\cite{McArdle2019,Yuan2019,Liu2021,Zhang2022}. On current hardware, these approaches prepare temperature-dependent states in small systems and approximate thermal states on larger spin lattices~\cite{Granet2026,Robertson2026,Farrell2026}. The larger-system demonstrations remain at relatively high effective temperatures, leaving low-temperature state preparation at scale out of reach. Reaching this regime requires shallow quantum circuits that reproduce the relevant correlations before accumulated hardware errors obscure them.

Where classical simulations remain tractable, they can help reduce the circuit resources needed for quantum state preparation. MPS representations can be compiled into quantum circuits, and MPS simulations can evaluate and optimize variational thermal-state circuits~\cite{Ran2020,Li2026}. Classical evaluations based on causal subcircuits and differentiable Pauli propagation provide further ways to determine circuit parameters before hardware execution~\cite{Drudis2026,Lin2026}. However, an accurate classical representation does not by itself ensure a shallow state preparation circuit. Using classical computation to construct the circuit directly makes its execution cost part of the design, alongside the accuracy of the prepared state.

In this work, we introduce PP-VarQITE, combining Pauli propagation~\cite{Rudolph2025} with variational quantum imaginary-time evolution (VarQITE)~\cite{McArdle2019} to determine shallow thermal-state preparation circuits classically. Starting from an infinite-temperature purification, we evaluate the metric matrix and evolution vector that determine the circuit's parameter updates through a shared Gram representation. Pauli propagation supplies the required circuit-dependent generator and Hamiltonian actions without constructing the full circuit state or propagating the thermal density matrix. The chosen circuit architecture therefore determines both the available state variations and the operators that must remain tractable in the classical calculation. As illustrated in Fig.~\ref{fig:workflow}, a single classical trajectory supplies parameters across target temperatures, while continued imaginary-time evolution changes the angles without adding gates to the state preparation circuit.

We numerically evaluate PP-VarQITE on one- and two-dimensional transverse-field Ising models (TFIM) and the frustrated $J_1$--$J_2$ Heisenberg chain, using up to 100 qubits including the purification register. Against exact solutions and independent thermal MPS references, we recover low-temperature energies and spatial ordering, including dimer formation in the frustrated chain and correlations extending across the Ising systems. For the frustrated chain and lattice, the variational trajectories recover this low-temperature behaviour after a delay relative to the thermal references at intermediate imaginary times. We obtain shallower state preparation circuits than the tested MPS compilation routes for both models, with an approximately twofold to tenfold reduction in depth for the frustrated chain. For the lattice, this depth advantage persists even where approximate MPS compilation gives smaller energy errors with fewer gates, identifying state preparation depth as a distinct resource in selecting circuits for hardware execution.

Using these shallow state preparation circuits, we prepare the 48-site frustrated chain and the $7\times7$ TFIM on Quantinuum Helios-1~\cite{Ransford2026} with 96 and 98 qubits, respectively. The measured energies and correlations resolve the development of dimer order throughout the chain and spatial order across the lattice. At target inverse temperatures $\beta J_1=6$ for the chain and $\beta|J|=1.5$ for the lattice, the hardware measurements retain the low-temperature ordering predicted for both models, with the lattice retaining approximately $94$--$96\%$ of the noiseless distance-averaged correlations. The compiled lattice circuits require at most 42 two-qubit layers, while the temperature dependence is encoded in the angles of a fixed state preparation architecture. These results establish a hardware benchmark for shallow thermal-state preparation across high- and low-temperature regimes in frustrated chains and two-dimensional lattices. The demonstrated state preparation depth motivates using these states to initialize dynamical calculations, with a larger share of the available circuit resources devoted to real-time evolution.

\section{Preliminaries}\label{sec:background}
We review thermal purification, variational imaginary-time evolution, and Pauli propagation to specify the ingredients used in our construction. Section~\ref{sec:method} then describes how we combine them into a classical compilation procedure.

\subsection{Thermal purification and the target trajectory}
We consider a Hamiltonian $H=\sum_{\mu=1}^{K}h_\mu P_\mu$ on $n$ physical qubits, with real coefficients $h_\mu$ and Hermitian Pauli strings $P_\mu$. We set $k_B=\hbar=1$. At inverse temperature $\beta$, our target is the Gibbs density matrix
\begin{equation}
 \rho_\beta=\frac{e^{-\beta H}}{Z_\beta},\qquad
 Z_\beta=\Tr(e^{-\beta H}),
 \label{eq:gibbs}
\end{equation}
where the partition function $Z_\beta$ ensures unit trace. Increasing $\beta$ reduces the relative populations of excited energy levels and connects the infinite-temperature state to the low-energy regime. In the limit $\beta\rightarrow\infty$, Eq.~\eqref{eq:gibbs} approaches the normalized projector onto the ground-state subspace. The physical reduced state is pure in this limit when the ground state is nondegenerate.

To represent the mixed state in Eq.~\eqref{eq:gibbs} with a unitary circuit, we introduce an $n$-qubit ancilla register. The circuit then acts on $N_q=2n$ qubits, and its pure output represents the physical density matrix after tracing out the ancillas. We initialize a product of physical--ancilla Bell pairs,
\begin{equation}
 \ket{\Phi}=2^{-n/2}\sum_{x=0}^{2^n-1}\ket{x}_S\ket{x}_A,
 \label{eq:bell}
\end{equation}
where $x$ labels computational basis strings and $S$ and $A$ denote the physical and ancilla registers. Tracing out $A$ leaves $\id_S/2^n$, so Eq.~\eqref{eq:bell} is a purification of the Gibbs state at $\beta=0$.

We obtain a finite-temperature purification by applying imaginary-time evolution to the physical register. With $\overline H=H\otimes\id_A$, the normalized state is
\begin{equation}
 \ket{\Psi_\beta}=\sqrt{\frac{2^n}{Z_\beta}}\, e^{-\beta\overline H/2}\ket{\Phi},\quad
 \Tr_A\ket{\Psi_\beta}\bra{\Psi_\beta}=\rho_\beta.
 \label{eq:purification}
\end{equation}
The factor of $1/2$ in the exponent of Eq.~\eqref{eq:purification} accounts for the imaginary-time operators acting on both the ket and the bra, which together produce $e^{-\beta H}$ in the reduced density matrix. We therefore evolve the purification in imaginary time $\tau=\beta/2$, with normalized evolution governed by
\begin{equation}
 \partial_\tau\ket{\Psi}=-(\overline H-E)\ket{\Psi},\quad E=\bra{\Psi}\overline H\ket{\Psi}.
 \label{eq:ite}
\end{equation}
Subtracting the instantaneous energy in Eq.~\eqref{eq:ite} preserves the norm while higher-energy components are suppressed. We derive both the normalization and this differential equation in Appendix~\ref{app:thermal}. 

\subsection{Variational quantum imaginary-time evolution}
In variational quantum imaginary-time evolution (VarQITE), we approximate the imaginary-time trajectory by $\ket{\psi(\thv)}=U(\thv)\ket{\mathbf{0}}$, where $\ket{\mathbf{0}}=\ket{0}^{\otimes N_q}$ and $U(\thv)$ includes all fixed gates, including Bell-pair preparation. The real parameter vector $\thv$ has $p$ components, each controlling a single rotation $R_i(\theta_i)=e^{-\mathrm{i}\theta_iG_i/2}$ generated by a Hermitian Pauli operator $G_i$.

Changing the parameters moves the state along the directions $\ket{\partial_i\psi}=\partial\ket\psi/\partial\theta_i$. By the chain rule, a parameter velocity $\dot{\thv}=d\thv/d\tau$ produces the state derivative
\begin{equation*}
 \frac{d}{d\tau}\ket{\psi(\thv(\tau))}
 =\sum_{i=1}^{p}\dot\theta_i\ket{\partial_i\psi}.
\end{equation*}
The circuit can follow the exact derivative in Eq.~\eqref{eq:ite} only through linear combinations of these available directions. Following McLachlan's principle~\cite{McArdle2019,Yuan2019}, we choose the real velocities that minimize the residual between the variational derivative and the desired imaginary-time derivative,
\begin{equation*}
 \min_{\boldsymbol{x}\in\mathbb R^p}
 \left\|\sum_i x_i\ket{\partial_i\psi}
 +(\overline H-E)\ket\psi\right\|^2.
\end{equation*}
Differentiating this quadratic expression with respect to each $x_i$ gives a linear system,
\begin{equation}
 M\dot{\thv}=\vv,
\label{eq:mclachlan}
\end{equation}
where
\begin{equation*}
 M_{ij}=\Rea\braket{\partial_i\psi}{\partial_j\psi}, \quad v_i=-\Rea\bra{\partial_i\psi}\overline H\ket\psi.
\end{equation*}
The metric $M$ contains the overlaps between the available state-derivative directions, while $\vv$ contains their real overlaps with the normalized imaginary-time derivative. Normalization implies $\Rea\braket{\partial_i\psi}{\psi}=0$, which removes the term proportional to $E$ from $v_i$. Determining the parameter velocities from Eq.~\eqref{eq:mclachlan} therefore requires evaluating these circuit-dependent overlaps at each imaginary-time step.

\subsection{Pauli propagation}

Pauli propagation evaluates circuit-dependent quantities by evolving operators in the Heisenberg picture while keeping the reference state fixed~\cite{Rall2019,Begusic2024,Rudolph2025}. For a circuit $U$ acting on $\ket{\mathbf{0}}$, we compute an expectation value as $\langle \mathbf{0}\rvert U^\dagger O U\lvert \mathbf{0}\rangle$ by expanding $O=\sum_P c_P P$ in Pauli strings and conjugating each string gate by gate. A Clifford gate maps each string to a signed Pauli string, whereas a rotation generated by a Pauli string can produce a linear combination. For $R=e^{-\mathrm{i}\theta G/2}$, where $G$ is a Hermitian Pauli string, the conjugation rule is

\begin{equation}
 R^\dagger P R=
 \begin{cases}
 P, & [G,P]=0,\\
 \cos\theta\,P+\mathrm{i}\sin\theta\,GP, & \{G,P\}=0.
 \end{cases}
 \label{eq:rotation}
\end{equation}
After propagation, the expectation value is obtained from strings containing only identities and $Z$ operators, since every string containing an $X$ or $Y$ factor has zero expectation value in $\ket{\mathbf 0}$.

An anticommuting string branches into two terms under Eq.~\eqref{eq:rotation}, so repeated branching can generate exponentially many terms in the number of non-Clifford rotations. Identical strings can be merged by summing their coefficients, but exact propagation can still become computationally prohibitive. To limit the representation size, one can discard terms according to their coefficient magnitude or their Pauli weight, defined as the number of nonidentity factors in a string. We specify the truncation procedure used for PP-VarQITE in Sec.~\ref{sec:method} and assess its effect on accuracy and classical cost in Sec.~\ref{sec:numerics}.

\section{PP-VarQITE}\label{sec:method}
We determine the parameters of a thermal-state preparation circuit by evaluating the variational quantum imaginary-time evolution (VarQITE) update with truncated Pauli propagation. The input is a Hamiltonian and a circuit architecture initialized to the Bell-pair purification. At each imaginary-time step, we evaluate a shared set of backward-propagated operators, assemble the metric and evolution vector, and update the parameters classically. The output is a set of circuit angles for the selected temperatures, as illustrated in Fig.~\ref{fig:workflow}.

\subsection{Shared Gram representation}
To evaluate the coefficients of Eq.~\eqref{eq:mclachlan} classically, we first separate the circuit around the rotation controlled by $\theta_i$ as $U=U_{>i}R_iU_{<i}$. We propagate the generator $G_i$ backward through the preceding gates and define
\begin{equation*}
 A_i=U_{<i}^{\dagger}G_iU_{<i},\quad \ket{a_i}=A_i\ket{\mathbf{0}},
\end{equation*}
\begin{equation}
 C=[\ket{a_1}\ \cdots\ \ket{a_p}].
\end{equation}
Because $G_i$ commutes with its own rotation, the derivative of the full circuit can be written as
\begin{equation*}
 \ket{\partial_i\psi}
 =-\frac{\mathrm{i}}2 U_{>i}R_iG_iU_{<i}\ket{\mathbf{0}}
 =-\frac{\mathrm{i}}2 U\ket{a_i}.
\end{equation*}
Thus, every derivative has the same final unitary $U$, and the vectors $\ket{a_i}$ express the different derivative directions in a common reference frame. The factor $U$ cancels from inner products between derivatives. We express the Hamiltonian in the same frame by defining $B=U^\dagger\overline H U$ and $\ket b=B\ket{\mathbf{0}}$. Substituting these expressions into Eq.~\eqref{eq:mclachlan} gives
\begin{equation}
 M=\frac14\Rea(C^\dagger C),\qquad
 \vv=\frac12\Ima(C^\dagger b).
 \label{eq:gram}
\end{equation}
Then, the energy can be written as $E=\Rea\braket{\mathbf{0}}{b}$. Equation~\eqref{eq:gram} is a Gram representation because its matrix entries are pairwise inner products of the columns of $C$. We propagate each generator once per coefficient evaluation and reuse its vector across the metric and evolution vector.

For any real vector $\boldsymbol z$, Eq.~\eqref{eq:gram} gives $\boldsymbol z^{\mathsf T}M\boldsymbol z=\|C\boldsymbol z\|^2/4\geq0$. The metric is therefore positive semidefinite, including when it is assembled from a shared set of truncated columns. Redundant derivative directions can still give zero eigenvalues, which we treat when solving the linear system. We derive the Gram expressions step by step and explain the unprojected metric convention used here in Appendix~\ref{app:pp-varqite-th}.

\subsection{Evaluating the metric and evolution vector with Pauli propagation}

We evaluate the generator actions by propagating each $G_i$ backward through its circuit prefix $U_{<i}$, and the Hamiltonian action by propagating each weighted term $h_\mu P_\mu\otimes\id_A$ through the full circuit $U$. After each gate, we merge identical Pauli strings and discard terms whose coefficient magnitude falls below $c_{\min}$, using the same threshold for generators and weighted Hamiltonian terms. We truncate the Hamiltonian terms independently before summing their actions, since combining terms before thresholding can change which coefficients are retained.

These truncated expansions give approximations $\ket{\widetilde a_i}$ and $\ket{\widetilde b}$ to the reference-state actions defined above. Each retained Pauli string maps $\ket{\mathbf 0}$ to one computational basis string with a known phase. We sum contributions to the same basis string and store only the nonzero amplitudes, avoiding dense vectors of dimension $2^{N_q}$ when the retained supports remain small. Inner products then involve only basis strings shared by the two vectors, with the amplitude assembly detailed in Appendix~\ref{app:pp-varqite-th}.

We assemble the approximate metric and evolution vector from these sparse overlaps,
\begin{equation*}
 \widetilde M_{ij}=\frac14\Rea
 \braket{\widetilde a_i}{\widetilde a_j},\qquad
 \widetilde v_i=\frac12\Ima
 \braket{\widetilde a_i}{\widetilde b}.
\end{equation*}
Reusing each generator action across the metric and evolution vector requires only $p$ generator and $K$ Hamiltonian-term propagations per evaluation. The subsequent assembly uses $p(p+1)/2$ inner products for the symmetric metric and $p$ for the evolution vector. The Hamiltonian action also supplies the energy estimate $\widetilde E=\Rea\braket{\mathbf 0}{\widetilde b}$.

To obtain the parameter velocities, we solve the approximate variational equations using Tikhonov-regularized least squares, reducing sensitivity to nearly dependent derivative directions. Starting from $\thv=\boldsymbol 0$, we advance the angles by $\Delta\tau\,\dot{\thv}$ using explicit Euler and save the parameter vector after each step. Successive saved vectors specify circuits at target inverse temperatures separated by $\Delta\beta=2\Delta\tau$, with all updates computed classically without hardware measurements. The regularized solve and integration settings are specified in Appendix~\ref{app:ablations}.

\subsection{Quantum circuit architectures}
We design a separate ansatz for the one-dimensional TFIM, the frustrated $J_1$--$J_2$ chain, and the two-dimensional TFIM, tailoring the generators and gate ordering to each model's interactions and expected correlations. The ansatz determines the state variations available during imaginary-time evolution, while its gate structure affects both state preparation depth and the classical cost of Pauli propagation. The accuracy and efficiency of PP-VarQITE therefore depend on this circuit design as well as on the approximations used to evaluate and integrate the variational evolution.

All three circuits prepare the exact infinite-temperature purification at zero angles, providing the initial state for the variational trajectory. Because the Hamiltonians and initial purification are real in the computational basis, exact imaginary-time evolution preserves real amplitudes, and we construct our ans\"atze to preserve this property. For the Ising models, we construct the circuits to include the interaction and transverse-field contributions to the initial imaginary-time derivative. For the frustrated chain, we choose the rotation ordering to accommodate the alternating singlet pattern while allowing correlations between neighboring dimers. We develop these designs through numerical testing and keep each architecture fixed throughout imaginary-time evolution, updating only its angles. Appendix~\ref{app:ansatz} gives the design rationale and complete gate sequences.

\section{Numerical results}\label{sec:numerics}

We evaluate how accurately PP-VarQITE reproduces thermal observables as system size increases, and determine the classical and quantum resources required. For each choice of ansatz and numerical settings, we integrate a single trajectory from zero angles, without restarts or separate optimization at each temperature. Unless stated otherwise, we use $\Delta\tau=0.05$, corresponding to $\Delta\beta=0.1$, and save the angles after every step. Saving the angles therefore provides state preparation circuits at successive target temperatures without changing the circuit architecture.

To control the classical propagation cost, we discard Pauli terms whose coefficient magnitude falls below $c_{\min}$, without imposing a Pauli-weight cutoff. We assess the resulting cost--accuracy tradeoff by varying this threshold and comparing the variational trajectories with independent thermal references. We also vary the number of circuit layers and the imaginary-time step to assess their effects on accuracy, with the corresponding convergence studies reported in Appendix~\ref{app:extended-numerics}.

\subsection{Models}

We choose the one-dimensional transverse-field Ising model (TFIM), the frustrated $J_1$--$J_2$ chain, and the two-dimensional TFIM as benchmarks of ordering and frustration, prioritizing independent classical validation over classical computational hardness. For each observable and system size, we use exact references where available and independently computed thermal matrix product states (MPS) references otherwise, with convergence checks reported in the appendix. These comparisons allow us to distinguish state preparation error from thermal excitation and assess the accuracy and circuit resources of PP-VarQITE across system sizes and temperatures. All calculations use Pauli operators and open boundary conditions (OBC).

For the one-dimensional TFIM,
\begin{equation*}
 H_{\mathrm{1D}}=J\sum_{i=1}^{n-1}Z_iZ_{i+1}+h_x\sum_{i=1}^{n}X_i,
\end{equation*}
we set $J=-1$ and $h_x=-0.5$. The free-fermion solution provides thermal energies and longitudinal correlations at every tested size~\cite{Pfeuty1970}.

The frustrated Heisenberg chain has
\begin{equation*}
 H_{J_1J_2}=J_1\sum_{i=1}^{n-1}\boldsymbol\sigma_i\cdot\boldsymbol\sigma_{i+1} + J_2\sum_{i=1}^{n-2}\boldsymbol\sigma_i\cdot\boldsymbol\sigma_{i+2},
\end{equation*}
where $\boldsymbol\sigma_i\cdot\boldsymbol\sigma_j=X_iX_j+Y_iY_j+Z_iZ_j$. We choose $J_1=1$ and $J_2=0.5$, where the even OBC chain has an exact singlet-dimer ground state with $E_0=-3J_1n/2$~\cite{Majumdar1969}. We assess the finite-temperature trajectory against thermal MPS references and use the exact ground energy to quantify the excess energy remaining at large inverse temperatures.

For the square lattice, we use
\begin{equation*}
 H_{\mathrm{2D}}=J\sum_{\langle i,j\rangle}Z_iZ_j+h_x\sum_iX_i,
\end{equation*}
with the same coefficients as the one-dimensional TFIM. 

\subsection{Validation strategy}

We assess state preparation accuracy through energies and spin correlations, using exact references where available and independent thermal MPS calculations otherwise. These observables allow validation at system sizes beyond practical full statevector simulation, with energies testing the variational trajectory and correlations testing the spatial ordering of the prepared states. Appendix~\ref{app:validation} describes the exact and approximate references for each model, while Appendix~\ref{app:mpsthermal} gives the MPS construction and convergence checks.

For the TFIM, we report longitudinal two-point correlators,
\begin{equation*}
 C_{ij}=\langle Z_iZ_j\rangle.
\end{equation*}
In the chain, we report correlations from the first site as $C(r)=C_{1,1+r}$. The spatial dependence of $C_{ij}$ tests whether the prepared states reproduce the growth of correlations across the system as the target temperature decreases.

For the frustrated Heisenberg chain, we assess singlet formation through the chain-averaged dimerisation,
\begin{equation*}
 b_i=\langle\boldsymbol\sigma_i\cdot\boldsymbol\sigma_{i+1}\rangle,\quad
 D=\bar b_{\mathrm{even}}-\bar b_{\mathrm{odd}},
\end{equation*}
where $\bar b_{\mathrm{even}}$ and $\bar b_{\mathrm{odd}}$ are the arithmetic means over even- and odd-indexed nearest-neighbor bonds, respectively. At infinite temperature, both means vanish and $D=0$. The exact singlet product on odd bonds has $\bar b_{\mathrm{odd}}=-3$ and $\bar b_{\mathrm{even}}=0$, giving $D=3$. Tracking $D$ therefore tests the development of dimer order toward the ground-state limit.

\subsection{Imaginary time evolution at scale}

\begin{figure*}[t]
 \centering
 \includegraphics[width=\textwidth]{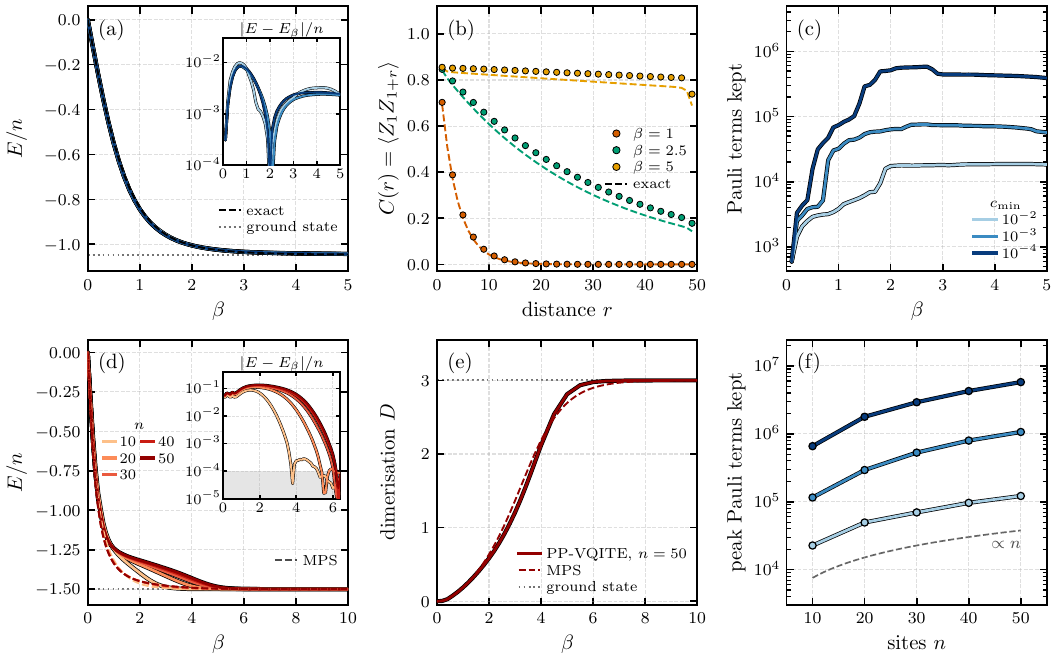}
 \caption{\textbf{Variational quantum imaginary-time evolution up to 100 qubits.}
The upper row shows the 50-site transverse-field Ising model (TFIM) with $J=-1$ and $h_x=-0.5$.
(a) Energy density $E/n$ and absolute thermal error $|E-E_\beta|/n$ in the inset vs. inverse-temperature $\beta$.
(b) Correlations $C(r)=\langle Z_1Z_{1+r}\rangle$ at $c_{\min}=10^{-4}$, with orange, green, and gold markers for $\beta=1,2.5,5$. Markers show only odd distances.
(c) Retained Pauli terms summed over generator and Hamiltonian-term propagation at each step.
Light to dark blue in (a,c) denotes $c_{\min}=10^{-2},10^{-3},10^{-4}$.
The lower row shows the frustrated chain with $J_1=1$ and $J_2=0.5$.
(d) Energy density for $n=10,20,30,40,50$, from light orange to dark red, at $c_{\min}=10^{-4}$.
The inset shows $|E-E_\beta|/n$, with gray shading below the approximate Pauli propagation resolution of $10^{-4}$ per site.
(e) Chain-averaged dimerisation $D=\bar b_{\mathrm{even}}-\bar b_{\mathrm{odd}}$ at $n=50$, where $b_i=\langle\boldsymbol\sigma_i\cdot\boldsymbol\sigma_{i+1}\rangle$.
(f) Peak retained-term counts for the three blue-coded cutoffs, with a dashed guide proportional to $n$.
Dashed curves in (a,b,d,e) give exact TFIM or thermal matrix product state (MPS) ($\chi=128$) references, and dotted lines mark ground-state values.}
 \label{fig:chains}
\end{figure*}

We investigate how PP-VarQITE follows the imaginary-time trajectory as the Ising and frustrated Heisenberg chains increase in size, using up to 100 qubits including the purification register. We use $L=2$ for the chain calculations presented here, with layer and time-step studies reported in Appendix~\ref{app:extended-numerics}.

We first compare the 50-site TFIM with exact imaginary-time evolution. The energy trajectory in Fig.~\ref{fig:chains}(a) remains close to the exact reference even at the coarsest propagation cutoff, $c_{\min}=10^{-2}$. Tightening the cutoff increases the retained Pauli representation but produces little change in the energy curve. We therefore recover the low-temperature energy for this model and circuit without retaining the smallest propagated coefficients.

The correlations in Fig.~\ref{fig:chains}(b) show that this energy agreement is accompanied by the expected spatial ordering. Correlators with $r>1$ are absent from the Hamiltonian and are evaluated after the parameter trajectory has been obtained, providing a test of spatial structure beyond the local terms that determine the energy. We reproduce the development of correlations spanning the chain as the target temperature decreases, with small positive deviations from the exact reference at larger separations in the low-temperature regime.

The retained-term count peaks at an intermediate imaginary time and decreases as the trajectory approaches the low-temperature regime, as shown in Fig.~\ref{fig:chains}(c). This nonmonotonic behaviour is consistent with a more demanding operator representation through the thermal crossover, followed by a simpler representation as spatial order develops.

For the frustrated chain, we recover low-temperature energies for systems from 10 to 50 physical sites after an intermediate-time lag relative to the MPS reference, as shown in Fig.~\ref{fig:chains}(d). The energy-density discrepancy increases with system size, but its growth weakens between the largest chains, as does the increase in imaginary time required for recovery. At 50 physical sites, the chain-averaged dimerisation in Fig.~\ref{fig:chains}(e) follows the development of singlet order toward $D=3$. The lag therefore delays the onset of the low-temperature regime along the variational trajectory, while the energy and dimerisation demonstrate recovery of the expected physics on 100 qubits.

We obtain these energy and dimerisation results with $c_{\min}=10^{-4}$, for which Fig.~\ref{fig:chains}(f) shows the growth of the peak retained-term count with system size. The 100-qubit trajectory reaches $\beta=10$ in about 1.8 minutes of evolution-step time on one classical node, excluding initialization and subsequent observable evaluation. The timing protocol is given in Appendix~\ref{app:ablations}, with convergence and scaling studies in Appendix~\ref{app:extended-numerics}. We choose the upper system size to match the scale of available quantum hardware, and the measured runtimes and retained-term counts suggest that the classical calculations can extend to larger systems.

\subsection{From classical representation to quantum state preparation}\label{sec:resources}

\begin{figure}[tbp]
 \centering
 \includegraphics[width=\columnwidth]{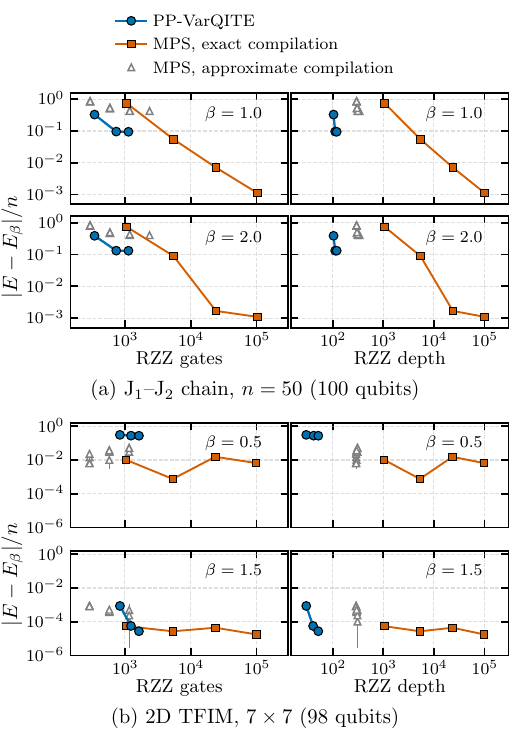}
 \caption{\textbf{Energy accuracy versus circuit gate count and depth.}
Absolute thermal energy-density error $|E-E_\beta|/n$ is plotted against two-qubit gate count on the left and depth on the right.
(a) The 50-site $J_1$--$J_2$ chain at $\beta=1,2$.
(b) The $7\times7$ transverse-field Ising model (TFIM) at $\beta=0.5,1.5$.
Blue circles show PP-VarQITE circuits with $L=1,2,3$ for the $J_1$--$J_2$ chain and $L=2,3,4$ for the lattice.
Orange squares show exact matrix product state (MPS) compilation at increasing bond dimension, and gray open triangles show individual randomized approximate MPS compilations with one, two, four, or eight staircase layers.
MPS references use $\chi=128$ for the $J_1$--$J_2$ chain and 256 for the lattice.}
 \label{fig:resources}
\end{figure}

Having obtained circuits at sizes beyond practical full statevector simulation, we assess whether their gate counts and depths are compatible with execution on noisy quantum hardware. We target Quantinuum Helios-1, which provides all-to-all connectivity and native single-qubit rotations and parametrized $R_{ZZ}$ gates~\cite{Ransford2026}. We therefore count $R_{ZZ}$ operations and measure two-qubit depth in layers of gates acting on disjoint qubits.

For MPS-based state preparation, exact compilation preserves the source state but produces multiqubit unitaries whose decomposition becomes costly as the bond dimension increases~\cite{Lin2021}. We therefore compress the reference MPS to selected bond dimensions and compare exact compilation of these sources with approximate compilation using a limited number of sequential sweeps of two-qubit gates~\cite{Ran2020}. The approximate route reduces circuit resources at the cost of additional error beyond the initial MPS compression. Compression and compilation procedures are detailed in Appendix~\ref{app:mpscompile}.

PP-VarQITE instead determines the angles of a prescribed circuit, allowing us to choose gate orderings that permit parallel rotations. For the $L=2$ chain circuits defined in Appendix~\ref{app:ansatz}, the two-qubit depths are $n+8$ for the one-dimensional TFIM and $2n+12$ for the $J_1$--$J_2$ chain. Both circuits use $2n$ qubits, so their qubit counts and state preparation depths grow linearly with the number of physical sites. We compare these directly parametrized circuits with the MPS compilation routes through their gate counts and depths at the achieved thermal energy errors in Fig.~\ref{fig:resources}, using the compilation settings in Appendices~\ref{app:resource-conventions} and~\ref{app:circuit-resources}.

For the $J_1$--$J_2$ chain on 100 qubits, the PP-VarQITE circuits in Fig.~\ref{fig:resources}(a) are approximately 2.5--3 times shallower than the shallowest tested MPS circuits and yield smaller energy errors than the approximate MPS compilations. Exact compilation at larger bond dimensions improves energy accuracy further, but requires orders of magnitude more gates and depth.

The depth advantage extends to the $7\times7$ TFIM on 98 qubits, where the PP-VarQITE circuits are approximately 6--10 times shallower than the shallowest tested MPS circuits (see Fig.~\ref{fig:resources}(b)). At $\beta=0.5$, approximate MPS compilation gives smaller energy errors with fewer gates, while by $\beta=1.5$ the PP-VarQITE energies approach the MPS reference.

By reducing state preparation depth at these system sizes, PP-VarQITE makes hardware execution accessible even where MPS compilation uses fewer gates. In the next section, we execute the circuits on Helios-1 and test whether their predicted energies and spatial correlations are retained in the presence of device noise.

\begin{figure}[tbp]
 \centering
 \includegraphics[width=\columnwidth]{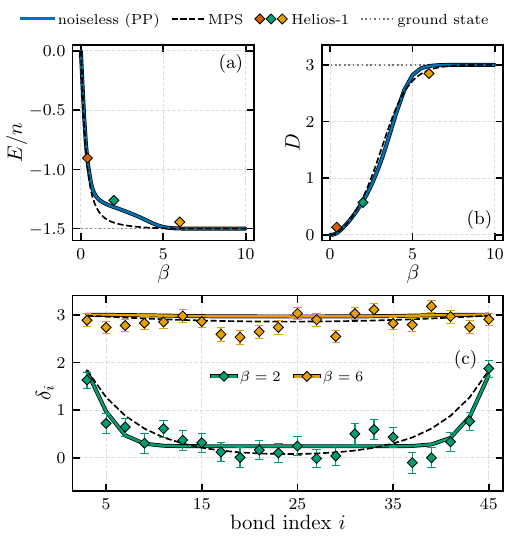}
\caption{\textbf{Frustrated-chain thermal state preparation on hardware.}
(a) Energy density $E/n$ vs inverse temperature $\beta$.
(b) Chain-averaged dimerisation $D=\bar b_{\mathrm{even}}-\bar b_{\mathrm{odd}}$, where $b_i=\langle\boldsymbol\sigma_i\cdot\boldsymbol\sigma_{i+1}\rangle$ vs. $\beta$.
(c) Local dimerisation $\delta_i=(b_{i-1}+b_{i+1})/2-b_i$ on interior odd bonds $i=3,5,\ldots,45$ at $\beta=2,6$.
Solid lines give noiseless circuit predictions estimated via Pauli propagation (PP), blue in (a,b) and coloured by inverse temperature in (c).
Black dashed lines give MPS references at $\chi=128$, and dotted lines mark the ground-state values in (a,b).
Diamonds show Helios-1 measurements at $\beta=0.4,2.0,6.0$, in vermillion, green, and gold, from 100 attempted shots in each of the $X$, $Y$, and $Z$ bases.
Error bars denote one sampling standard error and are smaller than the markers in (a,b).}
 \label{fig:hardware-j1j2}
\end{figure}

\section{Hardware demonstrations}\label{sec:hardware}

\begin{figure*}[tbp]
 \centering
 \includegraphics[width=\textwidth]{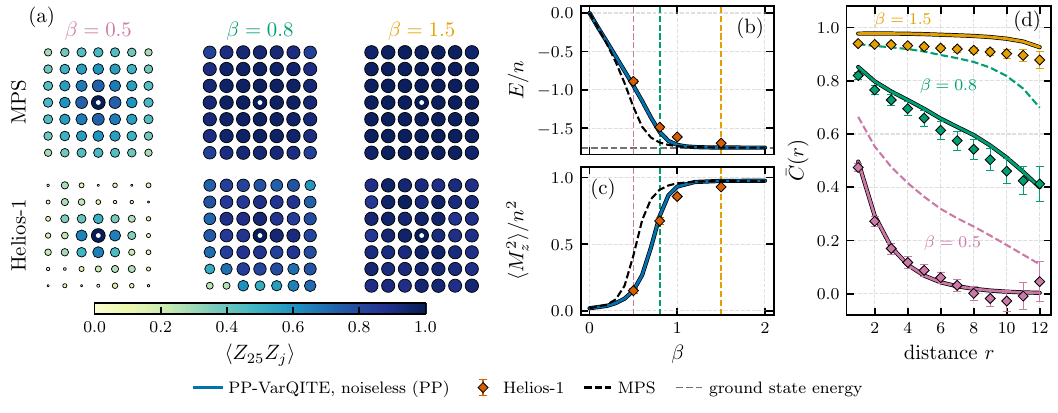}
 \caption{\textbf{Two-dimensional transverse-field Ising model (TFIM) thermal state preparation on hardware.}
Comparison of $7 \times 7$ TFIM thermal state preparation on hardware (98 qubits) to matrix product state (MPS) predictions.
(a) Spatial correlations with respect to the central site (site no. 25) $\langle Z_{25}Z_j\rangle$ at $\beta=0.5,0.8,1.5$, with MPS heatmaps above Helios-1 maps on a common colour scale.
(b) Energy density and (c) magnetic second moment $\langle M_z^2\rangle/n^2$  vs. inverse temperature $\beta$.
Blue solid curves give noiseless circuit predictions estimated via Pauli propagation (PP), black dashed curves give MPS references at $\chi=256$, and diamonds give hardware measurements at $\beta=0.5,0.8,1,1.5$.
Coloured vertical lines mark the map temperatures, and the gray dashed energy guide gives the second-order ground-state estimate.
(d) All pair correlations averaged at Manhattan distance $r$, with pink, green, and gold denoting $\beta=0.5,0.8,1.5$.
Solid curves give noiseless circuit predictions, dashed curves give MPS references, and diamonds give hardware estimates.
Hardware runs use 100 attempted shots in each of the $Z$ and $X$ bases.
Error bars denote one sampling standard error and can be smaller than the markers in panels (b,c,d).}
 \label{fig:hardware-2dtfim}
\end{figure*}

The shallow state preparation circuits obtained with PP-VarQITE enable us to prepare the 48-site frustrated chain and the $7\times7$ TFIM using 96 and 98 qubits on the Quantinuum Helios-1 trapped-ion quantum processor~\cite{Ransford2026}. The processor's all-to-all connectivity directly supports the couplings required by our ansätze, allowing their parallel gate layers to execute at native depth. In contrast, the compiled MPS circuits in Fig.~\ref{fig:resources} require hundreds of two-qubit layers, a depth at which accumulated hardware noise would obscure the target correlations. We determine all circuit angles classically with PP-VarQITE and execute the resulting state preparation circuits at selected imaginary times.

We assess the quality of state preparation on hardware by comparing the measured energies and correlations with noiseless predictions for the executed PP-VarQITE circuits and with thermal MPS references. These two baselines separate deviations caused by device noise from the variational error of the finite-depth ansatz along the imaginary-time trajectory. We compute the noiseless circuit predictions via truncated Pauli propagation without a hardware-noise model, using a truncation threshold for which classical approximation errors are negligible, as verified by the ablation studies in Appendix~\ref{app:extended-numerics}.

We first prepare the 48-site frustrated chain ($J_1=1$, $J_2=0.5$) on 96 qubits using the two-layer state preparation circuit obtained with PP-VarQITE, estimating the energy density and dimerisation from 100 attempted shots in each of the $X$, $Y$, and $Z$ measurement bases. As shown in Fig.~\ref{fig:hardware-j1j2}(a), the measured energy density tracks the decrease predicted along the variational trajectory from high to low temperatures, including its intermediate-imaginary-time lag relative to the thermal MPS reference. The accompanying growth of the chain-averaged dimerisation in Fig.~\ref{fig:hardware-j1j2}(b) demonstrates the formation of the singlet pattern across the chain. In the interior bond profiles of Fig.~\ref{fig:hardware-j1j2}(c), the hardware measurements at $\beta=2$ reproduce the stronger dimerisation near the boundaries and weaker dimerisation in the bulk, following the noiseless circuit prediction. By $\beta=6$, singlet order extends throughout the chain and the measured energy density remains close to the noiseless and thermal reference values. Tabulated energies and local dimerisation profiles across all measured temperatures are provided in Appendix~\ref{app:hardware-results}.

For the $7\times7$ TFIM ($J=-1$, $h_x=-0.5$), we execute the three-layer state preparation circuit obtained with PP-VarQITE on 98 qubits, requiring at most 42 two-qubit layers after compilation, and estimate the energy and longitudinal correlations from 100 attempted shots in each of the $X$ and $Z$ measurement bases. The central-site correlation heatmaps in Fig.~\ref{fig:hardware-2dtfim}(a) show the progression from short-range correlations to two-dimensional ferromagnetic order across the lattice. Comparing these maps, the energy density in Fig.~\ref{fig:hardware-2dtfim}(b), and the magnetic second moment in Fig.~\ref{fig:hardware-2dtfim}(c) with the MPS reference shows that the variational trajectory lags behind the thermal state at intermediate imaginary times before converging to the low-temperature reference by $\beta=1.5$. As shown by the distance-averaged correlations in Fig.~\ref{fig:hardware-2dtfim}(d), the hardware measurements at $\beta=0.5$ and $0.8$ closely follow the noiseless circuit predictions across Manhattan distance $r$, confirming that the departure from the MPS curve arises primarily from the variational lag rather than device noise. By $\beta=1.5$, the noiseless correlations converge to the thermal reference and the hardware measurements retain $94$--$96\%$ of the noiseless values across the lattice. Because evolving to larger imaginary times in PP-VarQITE updates only the angles of a fixed circuit, we recover this low-temperature order on the device without increasing the state preparation depth or hardware noise. Additional hardware data, including tabulated energies, corner-referenced heatmaps, and all-pairs distance averages, are provided in Appendix~\ref{app:hardware-results}.

Together, these 96- and 98-qubit demonstrations show that state preparation circuits obtained with PP-VarQITE remain accurate when executed directly on noisy quantum hardware. Without relying on deep compilation sequences or costly error mitigation, the classically optimized circuits prepare correlated thermal states that preserve local singlet pairing in the frustrated chain and long-range ferromagnetic order across the two-dimensional lattice from high to low temperatures. This performance extends PP-VarQITE from a classical method for estimating expectation values to a practical state preparation protocol that leaves most of a noisy processor's coherence budget available for subsequent quantum algorithms.

\section{Discussion}\label{sec:discussion}

In this work, we introduce PP-VarQITE to address a central obstacle to quantum thermal simulation on noisy processors, namely preparing correlated low-temperature states without deep circuits. By evaluating the McLachlan metric and evolution vector through a shared Gram representation, PP-VarQITE determines the rotation angles of a fixed-depth circuit classically via truncated Pauli propagation, encoding target temperatures along a single imaginary-time trajectory. Across one-dimensional frustrated and two-dimensional Ising benchmarks up to 100 qubits, this direct circuit parameterization reduces state preparation depth by roughly a factor of two to ten relative to tensor-network circuit compilation, enabling us to resolve low-temperature singlet pairing and lattice-wide ferromagnetic order on trapped-ion quantum hardware with up to 98 qubits. 

Propagating operators through a fixed-depth variational circuit avoids the low-temperature operator growth that limits direct Pauli propagation of thermal states. When imaginary-time evolution is applied directly to the thermal density matrix~\cite{GomezLurbe2026,Rudolph2026}, repeated operator multiplication causes the number of Pauli strings to grow rapidly as the temperature decreases. For the 30-site frustrated $J_1$--$J_2$ chain, this proliferation of Pauli terms prevented the density-matrix simulation of Rudolph et al.~\cite{Rudolph2026} from reaching the ground-state regime. In contrast, PP-VarQITE constrains operator branching by propagating the circuit generators and Hamiltonian terms through a shallow, fixed-depth ansatz. Instead of diverging at large $\beta$, the retained Pauli representation peaks near the thermal crossover, allowing our calculation to reach $\beta=10$ on the 30-site frustrated chain with at most $2.9\times10^6$ retained terms ($c_{\min}=10^{-4}$) and to scale to 50 physical sites while recovering the ground-state energy and dimer order. Propagating through a variational ansatz thus extends Pauli propagation to lower temperatures and larger systems while producing a shallow state preparation circuit for hardware execution.

Our benchmarks further show that an accurate classical representation of a thermal state does not by itself yield a shallow circuit for preparing it on hardware. For the models studied here, thermal MPS calculations provide accurate classical references, but compiling those representations into quantum circuits via sequential exact or staircase decompositions~\cite{Ran2020,Lin2021} introduces substantial circuit depth. By optimizing the angles of a parallelized circuit architecture directly, PP-VarQITE reduces the state preparation depth by roughly a factor of two to ten relative to the tested MPS compilations. Even at temperatures where approximate MPS compilation achieves a smaller energy error with fewer total gates, its sequential depth remains prohibitive for noisy hardware, whereas the shallower PP-VarQITE circuits execute accurately on the processor. While logarithmic-depth protocols for normal MPS~\cite{Malz2024} and further ansatz refinements may narrow this gap in one dimension, optimizing the target circuit architecture directly avoids the compilation bottleneck between a classical state representation and hardware execution.

Because PP-VarQITE projects imaginary-time evolution onto a pre-defined circuit architecture, the accuracy of the trajectory at intermediate temperatures is governed by the ansatz design. With the heuristic, model-tailored ans\"atze engineered here for the frustrated chain and two-dimensional lattice, the variational states lag behind the thermal reference at intermediate imaginary times before recovering the target ordering at larger $\beta$. When preparing states in this intermediate regime, evolving to a larger nominal imaginary time can compensate for the delay at fixed circuit depth, allowing an effective inverse temperature to be matched to reference observables. Eliminating the delay along the trajectory itself requires engineering ans\"atze whose generators better capture the imaginary-time derivative through the thermal crossover, motivating more systematic circuit constructions.

Constructing such circuits requires balancing the correlations an ansatz can represent against its state preparation depth and the classical cost of Pauli propagation. In our benchmarks, adding layers enlarges the variational family but increases operator branching and gate count without uniformly improving the integrated trajectory. Hamiltonian locality and symmetries provide a more systematic basis for selecting compact generators and reducing parameter counts in VarQITE~\cite{Wang2023}. Furthermore, unlike tensor-network methods such as MPS, which rely on a low-dimensional site ordering and incur rapidly growing bond dimensions in higher spatial dimensions, PP-VarQITE admits Hamiltonians and circuit ans\"atze with arbitrary interaction graphs and qubit connectivity. Extending thermal state preparation to higher-dimensional and nonlocally connected systems will therefore depend on designing shallow circuit generators that remain tractable under Pauli propagation.

The ability to prepare thermal states on flexible interaction graphs also addresses a central bottleneck in quantum machine learning, where energy-based generative models such as quantum Boltzmann machines (QBMs) represent target distributions through parametrized Gibbs states~\cite{Zoufal2021,Tuysuz2024QBM,Demidik2025Expressivity,Demidik2025SampleBased}. Although recent theoretical work establishes expressive power and polynomial sample-complexity guarantees for QBM training, these results assume efficient access to the model Gibbs state at each parameter update~\cite{Coopmans2024}. For QBM Hamiltonians that admit shallow ans\"atze and tractable Pauli propagation, PP-VarQITE offers a practical mechanism to supply these states by either estimating model expectation values classically during training or executing shallow circuits on hardware to draw samples from the learned distribution.

Beyond equilibrium sampling and generative modeling, shallow thermal state preparation circuits provide initial states for quantum simulation routines that evolve the system in real time. In finite-temperature dynamical simulations, initializing the processor in an approximate thermal state enables real-time evolution to extract dynamical correlation functions and response spectra~\cite{Sun2021,Lee2022}. In sample-based Krylov diagonalization, real-time evolutions of an initial reference state generate a subspace for classical diagonalization, with convergence and sampling overhead governed by the overlap between the initial state and the ground state~\cite{Yu2025}. Because increasing $\beta$ boosts this ground-state overlap and PP-VarQITE reaches larger $\beta$ by updating only the angles of a fixed shallow circuit, our framework can supply high-overlap initial states without deepening the state preparation stage. In both applications, thermal state preparation and subsequent real-time evolution share a limited coherence budget on noisy hardware. Offloading the state preparation stage to classical Pauli propagation, potentially in combination with Pauli-propagation-based compression of the subsequent real-time evolution~\cite{DAnna2025}, leaves a much larger share of the available circuit depth for resolving dynamical and spectral properties on the processor.

More broadly, our findings support treating classical operator propagation and quantum circuit design as a joint task to extend the reach of noisy quantum hardware. Together with recent demonstrations of classical Pauli-path optimization for ground-state circuits~\cite{LinGharibyan2026} and variational compression of two-dimensional dynamics~\cite{DAnna2025}, PP-VarQITE shows that classical simulation can produce compact, hardware-ready circuits that preserve many-body correlations on nearly 100 qubits across high- and low-temperature regimes. Developing systematic ansatz constructions that remain tractable under Pauli propagation could bring larger spin geometries and complete simulation workflows within the coherence budget of near-term quantum processors.

\medskip \noindent\textit{Note added.} While preparing this manuscript, we became aware of related work by Santona et al.~\cite{Santona2026}, who combined Majorana propagation with operator-projected variational imaginary-time evolution to train compact molecular ground-state circuits classically. Both approaches use classical operator propagation to determine variational imaginary-time parameter updates. In the present work, we use Pauli propagation to evaluate the state-derivative Gram metric for purified thermal states of spin models, and demonstrate shallow thermal state preparation circuits on noisy quantum hardware across a range of temperatures.

\begin{acknowledgments}
We thank Christa Zoufal and Marc Drudis for valuable discussions. We thank Reza Haghshenas and Sheng-Hsuan Lin for comments on the manuscript.
C.T. and M.G. are supported by CERN through the CERN Quantum Technology Initiative. This work is part of the Quantum Computing for High-Energy Physics (QC4HEP) working group.
\end{acknowledgments}

\section*{Declaration of AI use}
We conceived the core idea of PP-VarQITE independently of AI tools. We used LLMs to assist with refining the theoretical formulation, improving the implementation and its computational efficiency, and editing the manuscript. We take full responsibility for the correctness, integrity, and originality of the work, including all theoretical claims, numerical results, code, and manuscript content.

\section*{Data and code availability}
The data and code supporting this work are available from the authors upon reasonable request and will be made publicly available at a later date.

\onecolumngrid
\renewcommand{\bibsection}{\section*{References}}
\bibliographystyle{apsrev4-2}
\bibliography{references}

\clearpage
\setcounter{tocdepth}{2}
\begingroup
\footnotesize
\tableofcontents
\endgroup
\clearpage
\appendix

\section{Thermal purification and variational imaginary-time evolution}\label{app:thermal}

We derive the thermal purification convention and the McLachlan equations underlying the variational evolution. The derivation fixes the relation $\beta=2\tau$ between inverse temperature and pure-state imaginary time, and explains the normalization correction and projection onto circuit derivative directions. 

\subsection{Normalization and the imaginary-time convention}

For the Bell-pair purification considered here, evolving the physical register to imaginary time $\tau$ produces a Gibbs state at inverse temperature $\beta=2\tau$. We derive this factor of two and the corresponding norm-preserving evolution equation.

Let $H$ be a time-independent Hermitian Hamiltonian on a physical register $S$ of $n$ qubits, and let $A$ be an $n$-qubit ancilla register. The target Gibbs state is $\rho_\beta=e^{-\beta H}/Z_\beta$, where $\beta=(k_BT)^{-1}$ is the inverse temperature and $Z_\beta=\Tr(e^{-\beta H})$ is the partition function. Here, $T$ is the temperature and $k_B$ is Boltzmann's constant. We initialize the two registers in a product of $n$ Bell pairs,
\begin{equation}
 \ket{\Phi}
 =\frac{1}{\sqrt{2^n}}
 \sum_{x=0}^{2^n-1}\ket{x}_S\ket{x}_A,
 \label{eq:appbellstate}
\end{equation}
where $x$ labels $n$-bit computational basis states. We denote the identity operators on the registers by $\id_S$ and $\id_A$, with $\id_{SA}=\id_S\otimes\id_A$. Tracing out $A$ leaves the maximally mixed physical state $\id_S/2^n$, so $\ket\Phi$ is an infinite-temperature purification with $\beta=0$.

For an arbitrary physical-register operator $F$, expansion of the Bell-pair projector gives
\begin{equation}
\Tr_A\left[(F\otimes\id_A)\ket\Phi\bra\Phi
 (F^\dagger\otimes\id_A)\right]=\frac{1}{2^n}\sum_{x,y}
 F\ket{x}_S\bra{y}_S F^\dagger\,
 \Tr(\ket{x}_A\bra{y}_A)=\frac{1}{2^n}\sum_x
 F\ket{x}_S\bra{x}_S F^\dagger
 =\frac{FF^\dagger}{2^n}.
\label{eq:appbellidentity}
\end{equation}
The partial trace removes terms with $x\ne y$ because
$\Tr(\ket{x}_A\bra{y}_A)=\delta_{xy}$, and the final equality follows from
$\sum_x\ket{x}_S\bra{x}_S=\id_S$.

For imaginary time $\tau$, we choose $F=e^{-\tau H}$ and define the unnormalized purification
$\ket{u(\tau)}=e^{-\tau\overline H}\ket\Phi$, where
$\overline H=H\otimes\id_A$ acts only on the physical register.
Hermiticity of $H$ gives $FF^\dagger=e^{-2\tau H}$. The unnormalized reduced density matrix $\widetilde\rho_S(\tau)$ and the squared norm $\mathcal N(\tau)$ of the purification are therefore
\begin{equation}
\widetilde\rho_S(\tau)
 =\Tr_A\ket{u(\tau)}\bra{u(\tau)}
 =\frac{e^{-2\tau H}}{2^n},
\end{equation}%
\begin{equation}
 \mathcal N(\tau)
 =\braket{u(\tau)}{u(\tau)}
 =\Tr\left(\widetilde\rho_S(\tau)\right)
 =\frac{\Tr(e^{-2\tau H})}{2^n}
 =\frac{Z_{2\tau}}{2^n}.
\label{eq:appthermalnorm}
\end{equation}
The squared norm equals the trace of $\ket{u(\tau)}\bra{u(\tau)}$, which is unchanged by taking the partial trace over $A$.

Normalizing the purification by $\sqrt{\mathcal N(\tau)}$ divides its density matrix by $\mathcal N(\tau)$, giving
\begin{equation}
 \begin{aligned}
 \ket{\Psi(\tau)}
 &=\frac{\ket{u(\tau)}}{\sqrt{\mathcal N(\tau)}}
 =\sqrt{\frac{2^n}{Z_{2\tau}}}\,
 e^{-\tau\overline H}\ket\Phi,\\
 \rho_S(\tau)
 &=\Tr_A\ket{\Psi(\tau)}\bra{\Psi(\tau)}=\frac{\widetilde\rho_S(\tau)}{\mathcal N(\tau)}
 =\frac{e^{-2\tau H}}{Z_{2\tau}}.
 \end{aligned}
 \label{eq:appnormalizedpurification}
\end{equation}
Comparison with $\rho_\beta=e^{-\beta H}/Z_\beta$ gives $\beta=2\tau$. The factor of two arises because the density matrix contains one imaginary-time factor $e^{-\tau H}$ from the ket and another from the bra.

To obtain a differential equation for the normalized purification $\ket{\Psi(\tau)}$, we must differentiate both the unnormalized state $\ket{u(\tau)}$ and its normalization factor. Define the instantaneous energy
$E(\tau)=\bra{\Psi(\tau)}\overline H\ket{\Psi(\tau)}$.
Using $\partial_\tau\ket{u(\tau)}=-\overline H\ket{u(\tau)}$, we obtain
\begin{equation}
 \begin{aligned}
 \partial_\tau\mathcal N(\tau)
 &=\langle\partial_\tau u(\tau)\mid u(\tau)\rangle
 +\langle u(\tau)\mid\partial_\tau u(\tau)\rangle\\
 &=-2\bra{u(\tau)}\overline H\ket{u(\tau)}\\
 &=-2\mathcal N(\tau)E(\tau).
 \end{aligned}
 \label{eq:appnormderivative}
\end{equation}
Differentiating $\ket{\Psi(\tau)}=\mathcal N(\tau)^{-1/2}\ket{u(\tau)}$ then gives
\begin{equation}
 \begin{aligned}
 \partial_\tau\ket{\Psi(\tau)}
 &=\mathcal N(\tau)^{-1/2}\partial_\tau\ket{u(\tau)}
 -\frac{\partial_\tau\mathcal N(\tau)}
 {2\mathcal N(\tau)^{3/2}}\ket{u(\tau)}\\
 &=-\overline H\ket{\Psi(\tau)}
 -\frac{\partial_\tau\mathcal N(\tau)}
 {2\mathcal N(\tau)}\ket{\Psi(\tau)}\\
 &=-\left[\overline H-E(\tau)\right]\ket{\Psi(\tau)}.
 \end{aligned}
 \label{eq:appnormalizedite}
\end{equation}

The term $+E(\tau)\ket{\Psi(\tau)}$ compensates for the changing unnormalized norm $\mathcal N(\tau)$ by rescaling every component equally. To make its effect explicit, write
$\ket{\Psi(\tau)}=\sum_k c_k(\tau)\ket k$ in an eigenbasis of
$\overline H$, with $\overline H\ket k=\varepsilon_k\ket k$.
The amplitudes obey
$\partial_\tau c_k(\tau)=-[\varepsilon_k-E(\tau)]c_k(\tau)$, so
\begin{equation}
 \frac{c_k(\tau)}{c_\ell(\tau)}
 =\frac{c_k(0)}{c_\ell(0)}
 e^{-\tau(\varepsilon_k-\varepsilon_\ell)},
 \qquad c_\ell(0)\ne0.
 \label{eq:appamplituderatio}
\end{equation}
The common normalization factor cancels from this ratio. Higher-energy components are therefore suppressed relative to lower-energy components by the same factors as in unnormalized imaginary-time evolution, while the purification retains unit norm.

\subsection{Projection onto the circuit derivative directions}

The normalized imaginary-time derivative of a circuit state generally cannot be reproduced exactly by changing the circuit parameters. McLachlan's variational principle selects the parameter velocities that minimize the norm of this mismatch, giving the VarQITE update $M\dot{\thv}=\vv$.

Let $\ket{\psi(\thv)}$ be a normalized circuit purification with $p$ real parameters $\thv=(\theta_1,\ldots,\theta_p)$. We write
$\ket{\partial_j\psi}=\partial_{\theta_j}\ket{\psi(\thv)}$
for the state derivative with respect to $\theta_j$ and define the current energy
$E=\bra\psi\overline H\ket\psi$, where
$\overline H=H\otimes\id_A$.
Along imaginary time $\tau$, the chain rule gives
\begin{equation}
 \partial_\tau\ket{\psi(\thv(\tau))}
 =\sum_{j=1}^{p}\dot\theta_j\ket{\partial_j\psi},
 \qquad
 \dot\theta_j=\frac{d\theta_j}{d\tau}.
 \label{eq:appcircuitvelocity}
\end{equation}
Thus, changing the parameters generates state derivatives in the real span of the directions $\ket{\partial_j\psi}$.

At the same circuit state, the normalized imaginary-time equation prescribes the derivative
$-(\overline H-E)\ket\psi$.
For trial parameter velocities $\boldsymbol x\in\mathbb R^p$, the circuit derivative minus this target derivative defines the residual
\begin{equation}
 \ket{\mathcal R(\boldsymbol x)}
 =\sum_j x_j\ket{\partial_j\psi}
 +(\overline H-E)\ket\psi.
 \label{eq:appvarresidual}
\end{equation}
We choose the velocities to minimize
$\|\mathcal R(\boldsymbol x)\|^2$.
During this minimization, $\thv$ is fixed, so the current state $\ket\psi$, energy $E$, and derivative directions $\ket{\partial_j\psi}$ remain fixed while only $\boldsymbol x$ varies.

Since $\partial_{x_i}\ket{\mathcal R}=\ket{\partial_i\psi}$, differentiating the squared residual gives
\begin{equation}
 \frac{\partial\|\mathcal R(\boldsymbol x)\|^2}{\partial x_i}
 =\braket{\partial_i\psi}{\mathcal R(\boldsymbol x)}
 +\braket{\mathcal R(\boldsymbol x)}{\partial_i\psi}
 =2\Rea\braket{\partial_i\psi}{\mathcal R(\boldsymbol x)}.
 \label{eq:appstationarity}
\end{equation}
The squared residual is a convex quadratic function of the real velocities $\boldsymbol x$, so its minimizers satisfy these stationarity conditions. Setting the derivative to zero and substituting the residual gives
\begin{equation}
 \sum_j\Rea\braket{\partial_i\psi}{\partial_j\psi}\,x_j
 =-\Rea\bra{\partial_i\psi}\overline H\ket\psi
 +E\Rea\braket{\partial_i\psi}{\psi}.
 \label{eq:appnormalequations}
\end{equation}

The term proportional to $E$ vanishes because the circuit state remains normalized for every parameter choice. Differentiating $\braket{\psi}{\psi}=1$ with respect to $\theta_i$ gives
\begin{equation}
 0=\partial_{\theta_i}\braket{\psi}{\psi}
 =\braket{\partial_i\psi}{\psi}
 +\braket{\psi}{\partial_i\psi}
 =2\Rea\braket{\partial_i\psi}{\psi}.
 \label{eq:appnormalizationorthogonality}
\end{equation}
Identifying the remaining overlaps as
\begin{equation}
 M_{ij}=\Rea\braket{\partial_i\psi}{\partial_j\psi},
 \qquad
 v_i=-\Rea\bra{\partial_i\psi}\overline H\ket\psi,
 \label{eq:appmetricvector}
\end{equation}
the minimizing velocities $\boldsymbol x=\dot{\thv}$ satisfy
\begin{equation}
 M\dot{\thv}=\vv.
 \label{eq:appmclachlanupdate}
\end{equation}
This is the McLachlan equation used for the variational imaginary-time evolution.

If the derivative directions are linearly dependent over the real numbers, several parameter velocities generate the same state derivative and $M$ is singular. The projection then determines the closest available state derivative without uniquely determining the parameter velocities, so the treatment of redundant directions must be specified by the linear solver.

\clearpage
\section{PP-VarQITE formulation and classical evaluation}
\label{app:pp-varqite-th}

PP-VarQITE evaluates the McLachlan metric $M$ and evolution vector $\vv$ without constructing the full circuit state. We express the circuit derivatives and Hamiltonian action in a common input frame, then evaluate their overlaps through Pauli propagation. The exact identities are derived first, followed by the phase convention, sparse evaluation, and approximations used in the classical update.

\subsection{Backward-propagated generators and the Gram representation}

Let $\ket{\psi(\thv)}=U(\thv)\ket{\mathbf 0}$, where
$\ket{\mathbf 0}=\ket0^{\otimes N_q}$ and $N_q=2n$ includes the physical and ancilla registers. Each real parameter $\theta_i$ occurs in one Pauli rotation. Separating the circuit around that rotation gives
\begin{equation}
 U=U_{>i}R_iU_{<i},
 \qquad
 R_i=e^{-\mathrm{i}\theta_iG_i/2},
 \label{eq:appcircuitfactorization}
\end{equation}
where $G_i$ is a Hermitian Pauli string. Gates act from right to left, so $U_{<i}$ contains all gates applied before $R_i$, and $U_{>i}$ contains those applied after it.

To express each derivative in the circuit input frame, define the backward-propagated generator and its reference-state action,
\begin{equation}
 A_i=U_{<i}^\dagger G_iU_{<i},
 \qquad
 \ket{a_i}=A_i\ket{\mathbf 0}.
 \label{eq:appgeneratoractions}
\end{equation}
Since $G_i$ commutes with $R_i$, differentiation gives
\begin{equation}
 \begin{aligned}
 \ket{\partial_i\psi}
 &=-\frac{\mathrm{i}}2
 U_{>i}R_iG_iU_{<i}\ket{\mathbf 0}\\
 &=-\frac{\mathrm{i}}2
 U_{>i}R_iU_{<i}
 \left(U_{<i}^\dagger G_iU_{<i}\right)\ket{\mathbf 0}\\
 &=-\frac{\mathrm{i}}2 U\ket{a_i}.
 \end{aligned}
 \label{eq:appderivative}
\end{equation}
The second line inserts $U_{<i}U_{<i}^\dagger=\id$ between $R_i$ and $G_i$. Every derivative therefore contains the same final unitary $U$, which cancels from the derivative overlaps,
\begin{equation}
 \braket{\partial_i\psi}{\partial_j\psi}
 =\frac14\bra{a_i}U^\dagger U\ket{a_j}
 =\frac14\braket{a_i}{a_j}.
 \label{eq:appmetric}
\end{equation}

The Hamiltonian must be expressed in the same input frame. With
$\overline H=H\otimes\id_A$, define
\begin{equation}
 B=U^\dagger\overline H U,
 \qquad
 \ket b=B\ket{\mathbf 0}.
 \label{eq:apphamiltonianaction}
\end{equation}
Substitution into the evolution vector gives
\begin{equation}
 \begin{aligned}
 \bra{\partial_i\psi}\overline H\ket\psi
 &=\frac{\mathrm{i}}2
 \bra{a_i}U^\dagger\overline H U\ket{\mathbf 0}
 =\frac{\mathrm{i}}2\braket{a_i}{b},\\
 v_i
 &=-\Rea\left[\frac{\mathrm{i}}2\braket{a_i}{b}\right]
 =\frac12\Ima\braket{a_i}{b}.
 \end{aligned}
 \label{eq:appforce}
\end{equation}
The positive sign follows from $\Rea(\mathrm{i}z)=-\Ima z$. The factors $1/4$ in $M$ and $1/2$ in $\vv$ originate from the rotation convention in Eq.~\eqref{eq:appcircuitfactorization}.

Let $C$ contain the computational-basis amplitude columns of the states $\ket{a_i}$, and let $\boldsymbol b$ contain the amplitudes of $\ket b$. The coefficients then have the shared Gram representation
\begin{equation}
 M=\frac14\Rea(C^\dagger C),
 \qquad
 \vv=\frac12\Ima(C^\dagger\boldsymbol b).
 \label{eq:appgramcoefficients}
\end{equation}
Pauli propagation evaluates $A_i$ by conjugating $G_i$ backward through the circuit prefix $U_{<i}$. The Hamiltonian terms are conjugated through the full circuit $U$. Each generator action is evaluated once and reused across the metric and evolution vector.

In exact propagation, $A_i^\dagger=A_i$ and $A_i^2=\id$, so
$\braket{a_i}{a_i}=1$. Consequently,
\begin{equation}
 M_{ii}=\frac14,
 \qquad
 |M_{ij}|\leq\frac14,
 \label{eq:appdiagonal}
\end{equation}
where the off-diagonal bound follows from the Cauchy--Schwarz inequality.

\subsection{Phase convention for the real circuit ansätze}

Normalization requires $\Rea\braket{\psi}{\partial_i\psi}=0$, but the imaginary part can describe a parameter-dependent global phase. If an independent phase velocity is included in the McLachlan minimization, it cancels this component. Equivalently, the derivative directions are projected with
$Q=\id-\ket\psi\bra\psi$.

To express this projection in the input frame, define
$c_i=\bra{\mathbf 0}A_i\ket{\mathbf 0}\in\mathbb R$.
Equation~\eqref{eq:appderivative} gives
$\braket{\psi}{\partial_i\psi}=-\mathrm{i}c_i/2$, and the phase-projected metric is
\begin{equation}
 M^{\mathrm{proj}}_{ij}
 =\Rea\bra{\partial_i\psi}Q\ket{\partial_j\psi}
 =\frac14\Rea\left[
 \braket{a_i}{a_j}-c_ic_j\right].
 \label{eq:appphase}
\end{equation}
The evolution vector is unchanged because the normalized imaginary-time target is orthogonal to $\ket\psi$.

For the ansätze used in the parameter evolution, all fixed gates are real and each Pauli generator contains an odd number of $Y$ factors. Such a generator is purely imaginary, so its rotation $e^{-\mathrm{i}\theta_iG_i/2}$ is real. The circuit state and its parameter derivatives are therefore real. Normalization then implies
$\braket{\psi}{\partial_i\psi}=0$, giving $c_i=0$.
The projected and unprojected exact metrics agree for these ansätze, and we use the unprojected Gram representation.

\subsection{Sparse amplitudes and overlap evaluation}

The Gram representation requires the actions of $A_i$ and $B$ on the reference state, rather than dense matrices for these operators. For a Pauli expansion $O=\sum_P c_P P$, let $x(P)$ be the bit string with a one wherever $P$ contains $X$ or $Y$, and let $\nu_Y(P)$ count its $Y$ factors. The identities
$X\ket0=\ket1$, $Y\ket0=\mathrm{i}\ket1$, and $Z\ket0=\ket0$ give
\begin{equation}
 O\ket{\mathbf 0}
 =\sum_P c_P\,\mathrm{i}^{\nu_Y(P)}\ket{x(P)}.
 \label{eq:sparseaction}
\end{equation}
Several Pauli strings can contribute to the same basis amplitude. For example,
\begin{equation}
 \begin{aligned}
 O&=c_1X\otimes\id+c_2X\otimes Z+c_3Y\otimes\id,\\
 O\ket{00}&=(c_1+c_2+\mathrm{i}c_3)\ket{10}.
 \end{aligned}
 \label{eq:appsparseexample}
\end{equation}
The coefficients and phases must therefore be combined before taking inner products.

Writing $A_i=\sum_P c_{iP}P$ and $B=\sum_P d_PP$, the required amplitudes are
\begin{equation}
 C_{xi}=\sum_{P:\,x(P)=x}c_{iP}\mathrm{i}^{\nu_Y(P)},
 \qquad
 b_x=\sum_{P:\,x(P)=x}d_P\mathrm{i}^{\nu_Y(P)}.
 \label{eq:appsparseamplitudes}
\end{equation}
The overlaps are then evaluated as
\begin{equation}
 M_{ij}=\frac14\Rea\sum_x C_{xi}^*C_{xj},
 \qquad
 v_i=\frac12\Ima\sum_x C_{xi}^*b_x.
 \label{eq:appsparsesums}
\end{equation}
Only basis strings present in both vectors contribute to an inner product. We store the nonzero amplitudes indexed by their basis strings, so dense vectors of dimension $2^{N_q}$ are unnecessary when the supports remain small. The exact Hamiltonian action also gives
$E=\bra{\mathbf 0}B\ket{\mathbf 0}=b_{\mathbf 0}$, which is real because $B$ is Hermitian.

\subsection{Coefficient truncation and consistency of the Gram metric}

To limit classical propagation cost, we discard Pauli terms whose coefficient magnitude is below $c_{\min}$ after each gate. Identical strings are merged before applying the threshold, and no Pauli-weight cutoff is imposed. Denote this gate-by-gate approximation to $V^\dagger OV$ by
$\mathcal P_{c_{\min}}(O;V)$.

For $H=\sum_{\mu=1}^{K}h_\mu P_\mu$, the retained operators are
\begin{equation}
 \widetilde A_i
 =\mathcal P_{c_{\min}}(G_i;U_{<i}),
 \qquad
 \widetilde B
 =\sum_{\mu=1}^{K}
 \mathcal P_{c_{\min}}(h_\mu P_\mu\otimes\id_A;U).
 \label{eq:apptruncatedoperators}
\end{equation}
Each weighted Hamiltonian term is propagated independently, with the threshold applied to its full coefficient, including $h_\mu$. Truncating the sum of propagated terms would define a different approximation because coefficient thresholding is nonlinear.

Applying these operators to $\ket{\mathbf 0}$ gives the retained vectors
$\ket{\widetilde a_i}=\widetilde A_i\ket{\mathbf 0}$ and
$\ket{\widetilde b}=\widetilde B\ket{\mathbf 0}$.
Let $\widetilde C$ and $\widetilde{\boldsymbol b}$ contain their sparse amplitudes. We assemble
\begin{equation}
 \widetilde M=\frac14\Rea(\widetilde C^\dagger\widetilde C),
 \qquad
 \widetilde{\vv}
 =\frac12\Ima(\widetilde C^\dagger\widetilde{\boldsymbol b}).
 \label{eq:appapproximategram}
\end{equation}
After truncation, $\widetilde A_i^2=\id$ need not hold, so the retained columns need not have unit norm and $\widetilde M_{ii}$ need not equal $1/4$.

Positive semidefiniteness nevertheless follows from using the same retained columns for every metric entry. For any real vector $\boldsymbol z$,
\begin{equation}
 \boldsymbol z^{\mathsf T}\widetilde M\boldsymbol z
 =\frac14\Rea\left[
 (\widetilde C\boldsymbol z)^\dagger
 (\widetilde C\boldsymbol z)\right]
 =\frac14\|\widetilde C\boldsymbol z\|^2\geq0.
 \label{eq:apppsd}
\end{equation}
Replacing selected entries, such as restoring the diagonal to $1/4$, would define a different matrix and would not in general preserve this property. The shared-column construction preserves positive semidefiniteness, while agreement with the untruncated coefficients and physical observables requires convergence checks.

\clearpage
\section{Ansatz design}\label{app:ansatz}

We tailor the circuit generators and gate ordering to each model's interactions and expected correlations, while keeping the architecture fixed throughout each imaginary-time trajectory. We develop these ansätze through empirical iteration and numerical testing. The resulting designs are not necessarily optimal, and further refinement may improve state preparation accuracy or reduce circuit resources.

\subsection{Common conventions}

We denote the physical and ancillary qubits at site $i$ by $d_i$ and $a_i$, respectively, and order the wires as $d_1,a_1,d_2,a_2,\ldots$. Each parametrized gate has an independent angle, with
$R_P(\theta)=e^{-\mathrm{i}\theta P/2}$.
The Hamiltonians and initial Bell-pair state have real amplitudes, which exact imaginary-time evolution preserves. We therefore use Pauli generators containing an odd number of $Y$ factors, so that their rotations also preserve real amplitudes. At zero angles, all rotations are identities and the CNOT networks preceding Bell state preparation leave the computational zeros unchanged. Each circuit therefore starts from the exact infinite-temperature purification
$\ket\Phi=\bigotimes_{i=1}^n\ket{\Phi^+}_{d_i a_i}$,
where $\ket{\Phi^+}=(\ket{00}+\ket{11})/\sqrt{2}$.

\subsection{Ansatz design for the one-dimensional transverse-field Ising model}

\begin{figure}[htbp]
 \centering
 \includegraphics[width=0.58\textwidth]{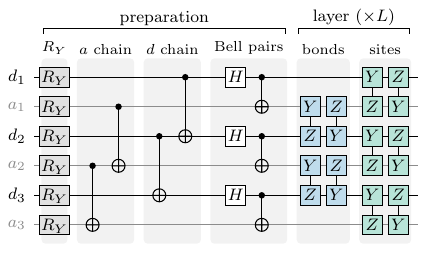}
 \caption{\textbf{Example of the Ising-chain state preparation circuit.} Wires are ordered $d_1,a_1,d_2,a_2,\ldots$, here for three sites and one layer. Each connected Pauli-letter group represents one rotation with its own angle. The initial rotations and descending CNOT networks precede Bell state preparation and the bond and site layers. Gray ancilla wires alternate with black data wires. Shaded stages and upper brackets identify state preparation and the repeated layer, with blue bond rotations and green site rotations. The drawing preserves gate order rather than an executed device schedule.}
 \label{fig:circuits}
\end{figure}

We design the Ising-chain circuit to supply both the interaction and transverse-field directions needed to begin imaginary-time evolution from infinite temperature. Because the Hamiltonian is traceless, the initial energy vanishes and the exact imaginary-time derivative is $-\overline H\ket\Phi$. The circuit tangent space must therefore contain $Z_{d_i}Z_{d_{i+1}}\ket\Phi$ for every interaction bond and $X_{d_i}\ket\Phi$ for every site. We obtain these directions through the initial state preparation stage and the local data--ancilla rotations in Fig.~\ref{fig:circuits}.

The initial data-qubit $R_Y$ rotations supply the interaction directions through the descending CNOT chain. A first-order excitation at site $i<n$ is copied to site $i+1$. Because the CNOTs farther along the chain have already acted, the excitation reaches only that neighboring site. Bell state preparation then converts the two data-qubit flips into the required $Z_{d_i}Z_{d_{i+1}}$ direction. The on-site rotation $R_{Y_{d_i}Z_{a_i}}$ supplies the transverse-field direction. Writing their angles as $\alpha_i$ and $\eta_i$, respectively, gives
\begin{equation}
 \left.\partial_{\alpha_i}\ket{\psi(\thv)}\right|_{\thv=\boldsymbol 0}
 =\frac12 Z_{d_i}Z_{d_{i+1}}\ket\Phi,
 \qquad
 \left.\partial_{\eta_i}\ket{\psi(\thv)}\right|_{\thv=\boldsymbol 0}
 =\frac12 X_{d_i}\ket\Phi.
 \label{eq:appinitialtfimdirections}
\end{equation}
Thus, the circuit contains the directions required for the initial imaginary-time derivative. The repeated bond and site rotations provide additional variational freedom as the state moves away from the Bell-pair initialization.

The complete gate sequence, in time order, is
\begin{enumerate}
 \item Apply $R_Y$ to all $2n$ qubits.
 \item Apply $\mathrm{CNOT}(a_i\to a_{i+1})$ for $i=n-1,\ldots,1$, followed by $\mathrm{CNOT}(d_i\to d_{i+1})$ in the same descending order.
 \item Prepare Bell pairs by applying $H(d_i)$ followed by $\mathrm{CNOT}(d_i\to a_i)$ at every site.
 \item Repeat $L$ layers. Each layer applies $R_{Y_{a_i}Z_{d_{i+1}}}$ and $R_{Z_{a_i}Y_{d_{i+1}}}$ for $i=1,\ldots,n-1$, followed by $R_{Y_{d_i}Z_{a_i}}$ and $R_{Z_{d_i}Y_{a_i}}$ for $i=1,\ldots,n$.
\end{enumerate}
The initial stage contributes $2n$ parameters, and each repeated layer contributes two rotations per bond and two per site, giving
\begin{equation}
 p_{\mathrm{TFIM\text{-}1D}}
 =2n+L\bigl[2(n-1)+2n\bigr]
 =2n+L(4n-2).
 \label{eq:appparam1d}
\end{equation}
We keep $L$ fixed for all values of $\beta$ and update only the rotation angles, so additional integration steps do not add gates to the state preparation circuit.

\subsection{Ansatz design for the one-dimensional \texorpdfstring{$J_1$--$J_2$}{J1-J2} model}\label{app:j1j2ansatz}

\begin{figure}[htbp]
 \centering
 \includegraphics[width=\textwidth]{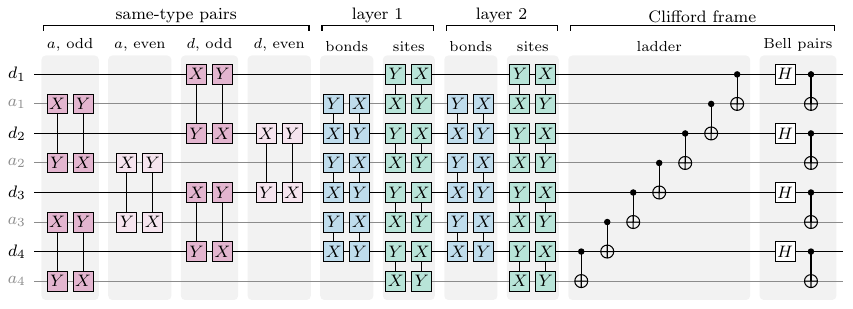}
 \caption{\textbf{Frustrated-chain circuit with odd-first grouping.} The example has four physical sites, eight purification qubits, and $L=2$. Each connected Pauli-letter group denotes one rotation with an independent angle. Purple rotations act within each register, with darker and lighter shading distinguishing odd and even bonds. Blue rotations couple neighboring data--ancilla sites, and green rotations couple the qubits within each site. All rotations precede the descending CNOT ladder and Bell state preparation. Black and gray wires denote data and ancillary qubits, respectively. Shaded bands and upper brackets identify the circuit stages.}
 \label{fig:j1j2circuit}
\end{figure}

We design the frustrated-chain circuit to accommodate the alternating singlet pattern of the ground state while allowing correlations between neighboring dimers. At $J_2/J_1=1/2$, the even chain has a singlet covering on $(1,2),(3,4),\ldots$. We use this covering to choose the order of the same-register rotations in Fig.~\ref{fig:j1j2circuit}, with odd bonds applied before even bonds. The odd-bond rotations act on disjoint pairs throughout the chain, while the subsequent even-bond rotations couple those pairs. Because rotations on overlapping bonds need not commute, this ordering is part of the variational ansatz.

All parametrized rotations precede a fixed Clifford frame consisting of a descending CNOT ladder followed by Bell state preparation. To explain this placement, write $V(\thv)$ for the parametrized gates and $C_J=C_\Phi C_{\mathrm{lad}}$ for the frame. Since $C_J\ket{\mathbf0}=\ket\Phi$, the prepared state can be written as
\begin{equation}
 \ket{\psi(\thv)}
 =C_JV(\thv)\ket{\mathbf0}
 =\bigl[C_JV(\thv)C_J^\dagger\bigr]\ket\Phi.
 \label{eq:appjframe}
\end{equation}
A rotation with generator $P$ before the frame therefore acts on the Bell-pair state through the transformed generator $C_JPC_J^\dagger$. This placement allows two-qubit rotations to produce variations involving several qubits in the purification while retaining two-qubit parametrized gates.

We use $XY$ and $YX$ rotations, which mix computational-basis states in pairs while preserving real amplitudes. Rotations within each register provide correlations between neighboring sites, while the subsequent data--ancilla blocks couple the registers both across neighboring sites and within each site. Every rotation has an independent angle, so the odd-first ordering favors a pairing pattern without fixing its bond strengths or imposing singlets during imaginary-time evolution.

The complete gate sequence, in time order, is
\begin{enumerate}
 \item Repeat $L-1$ same-register blocks. On the ancillas, apply $R_{X_{a_i}Y_{a_{i+1}}}$ followed by $R_{Y_{a_i}X_{a_{i+1}}}$ on odd bonds $i=1,3,\ldots$, then on even bonds $i=2,4,\ldots$. Apply the corresponding sequence on the data register, again with odd bonds before even bonds.
 \item Repeat $L$ data--ancilla blocks. Each block applies $R_{Y_{a_i}X_{d_{i+1}}}$ and $R_{X_{a_i}Y_{d_{i+1}}}$ for $i=1,\ldots,n-1$, followed by $R_{Y_{d_i}X_{a_i}}$ and $R_{X_{d_i}Y_{a_i}}$ for $i=1,\ldots,n$.
 \item Apply the descending ladder $\mathrm{CNOT}(q_k\to q_{k+1})$ for $k=2n-1,\ldots,1$, with $(q_1,q_2,\ldots)=(d_1,a_1,d_2,a_2,\ldots)$.
 \item Prepare Bell pairs by applying $H(d_i)$ followed by $\mathrm{CNOT}(d_i\to a_i)$ at every site.
\end{enumerate}
Each same-register block contributes $4(n-1)$ parameters, and each data--ancilla block contributes $4n-2$, giving
\begin{equation}
(L-1)(4n-4)+L(4n-2)
 \label{eq:appparamj1j2}
\end{equation}
parameters. Under the all-to-all connectivity assumption, the parametrized blocks have a constant depth per block, whereas the fixed CNOT ladder contributes a depth linear in $n$. 

\subsection{Ansatz design for the two-dimensional transverse-field Ising model}\label{app:tfim2dansatz}

\begin{figure}[htbp]
 \centering
 \begin{minipage}[c]{0.34\textwidth}
  \centering\textbf{(a)}\par\smallskip
  \includegraphics[width=\linewidth]{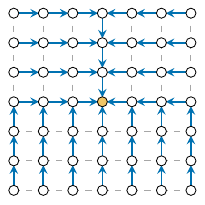}
 \end{minipage}\hfill
 \begin{minipage}[c]{0.60\textwidth}
  \centering\textbf{(b)}\par\smallskip
  \includegraphics[width=\linewidth]{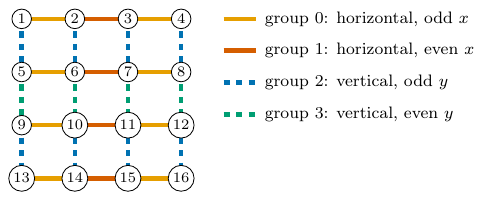}
 \end{minipage}
 \caption{\textbf{Spanning-tree state preparation and full-lattice bond rotations.}
 (a) The state preparation tree on a $7\times7$ lattice. Blue arrows denote data-qubit CNOTs directed toward the central orange root, with gates nearer the root applied first. Gray dashed bonds lie outside the tree.
 (b) The four groups used to apply bond rotations, illustrated on a $4\times4$ lattice. Orange and red solid bonds distinguish the two horizontal groups, while blue and teal dashed bonds distinguish the two vertical groups. Bonds within each group have disjoint sites. Dashed lines in the two panels have different meanings.}
 \label{fig:bondgroups}
\end{figure}

We extend the Ising-chain construction by preparing bond directions along a spanning tree and adding rotations for the interactions outside that tree. A spanning tree reaches every site using $n-1$ bonds, but the two-dimensional lattice also contains bonds that close loops. The initial state preparation therefore supplies only part of the interaction structure, as illustrated in Fig.~\ref{fig:bondgroups}(a). We include explicit bond rotations so that the circuit tangent space contains the directions required to begin imaginary-time evolution under the full lattice Hamiltonian.

The state preparation tree acts only on data qubits. Each CNOT points from a child $c$ to its parent $p$, with gates nearer the root applied first. As in the chain construction, an initial data-qubit rotation then produces a first-order excitation on the child and its parent. The excitation stops there because the parent's CNOT toward the root has already acted. Bell state preparation converts these two flips into the interaction direction $Z_{d_c}Z_{d_p}\ket\Phi$. The ancillas have no tree network, so their initial rotations independently supply the transverse-field directions $X_{d_i}\ket\Phi$.

To supply the interaction directions outside the tree, each variational layer includes $R_{Y_{d_i}X_{a_i}Z_{d_j}}$ on every lattice bond $(i,j)$. At the Bell-pair initialization, its derivative is
\begin{equation}
 -\frac{\mathrm{i}}{2}
 Y_{d_i}X_{a_i}Z_{d_j}\ket\Phi
 =-\frac12 Z_{d_i}Z_{d_j}\ket\Phi.
 \label{eq:app2dbonddirection}
\end{equation}
These rotations therefore provide the Ising-bond directions on the full lattice, including bonds omitted from the tree. The additional two-qubit bond and site rotations allow intersite and data--ancilla correlations to vary as imaginary-time evolution proceeds away from the initial state.

\begin{figure}[htbp]
 \centering
 \includegraphics[width=\textwidth]{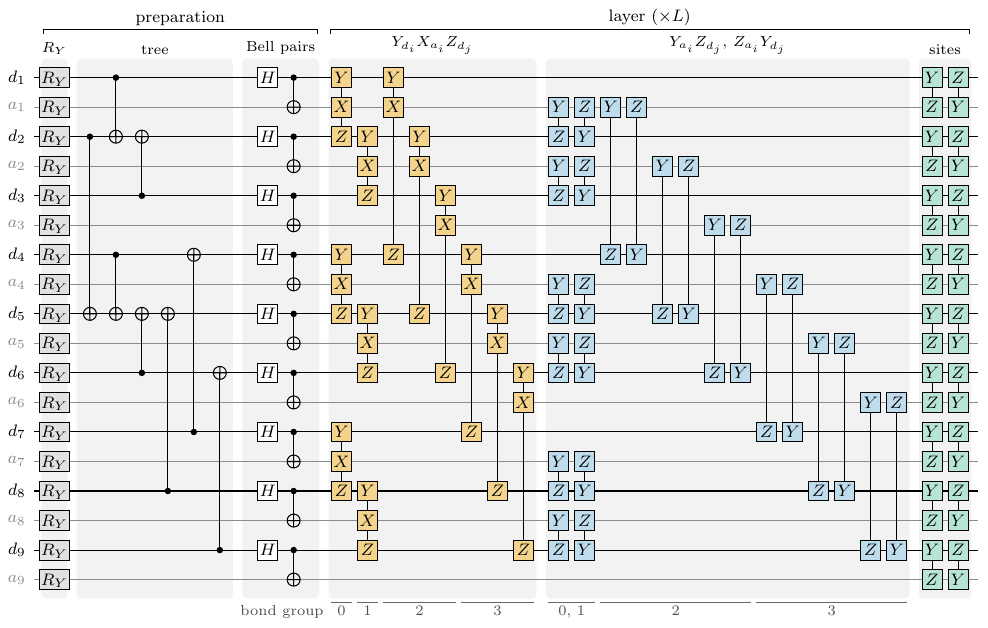}
 \caption{\textbf{Lattice circuit with tree state preparation and bond rotations.}
 The example has nine physical sites, eighteen purification qubits, and $L=1$. Black and gray wires denote data and ancillary qubits, respectively. Gray $R_Y$ gates precede the data-register tree and Bell state preparation. Orange connected Pauli-letter groups denote three-qubit bond rotations, blue groups denote two-qubit bond rotations, and green groups act within a site. Each connected group has an independent angle. Shaded stages and upper brackets distinguish state preparation from the repeated layer, while labels below the circuit identify the four bond groups.}
 \label{fig:tfim2dcircuit}
\end{figure}

For an $L_x\times L_y$ lattice, we label sites in row-major order as $i=(y-1)L_x+x$. We construct the tree by breadth-first search from the central site, visiting neighbors in increasing site-label order. This choice limits the longest root-to-site path while fixing the gate order needed to reproduce the circuit. The complete sequence, illustrated in Fig.~\ref{fig:tfim2dcircuit}, is
\begin{enumerate}
 \item Apply $R_Y$ to all $2n$ qubits.
 \item Apply $\mathrm{CNOT}(d_c\to d_p)$ on every tree edge in breadth-first order, beginning with edges nearest the root.
 \item Prepare Bell pairs by applying $H(d_i)$ followed by $\mathrm{CNOT}(d_i\to a_i)$ at every site.
 \item Repeat $L$ layers. First apply $R_{Y_{d_i}X_{a_i}Z_{d_j}}$ on every lattice bond $(i,j)$ with $i<j$. Next apply $R_{Y_{a_i}Z_{d_j}}$ and $R_{Z_{a_i}Y_{d_j}}$ on every bond. Finish with $R_{Y_{d_i}Z_{a_i}}$ and $R_{Z_{d_i}Y_{a_i}}$ at every site.
\end{enumerate}
In each bond stage, we apply horizontal bonds with odd left-endpoint $x$, horizontal bonds with even left-endpoint $x$, vertical bonds with odd upper-endpoint $y$, and vertical bonds with even upper-endpoint $y$, in that order. Each group contains disjoint site pairs, allowing parallel rotations as shown in Fig.~\ref{fig:bondgroups}(b). Rotations in different groups can overlap, so we retain the group order as part of the ansatz.

The initial rotations contribute $2n$ parameters, while each layer contributes three rotations per lattice bond and two per site. With $n=L_xL_y$ and $N_b=2n-L_x-L_y$, the total is
\begin{equation}
 p_{\mathrm{TFIM\text{-}2D}}
 =2n+L(3N_b+2n).
 \label{eq:appparam2d}
\end{equation}
The four bond groups give a size-independent depth per variational layer under the all-to-all connectivity assumption. The tree is applied only once during state preparation.

\clearpage
\section{Numerical implementation and validation details}\label{app:ablations}

\subsection{Software and hardware information}

The PP-VarQITE evolution uses Julia 1.12.5 with PauliPropagation.jl 0.4.1~\cite{Rudolph2025}. Sparse reference-state actions use fixed-width bit strings, with UInt128 keys for up to 128 circuit qubits. The MPS reference environment uses Python 3.14.3, quimb 1.13.0, NumPy 2.4.4, SciPy 1.17.1 and \texttt{mps-to-circuit} 0.1.2~\cite{MPStoCircuit2025}.  We export circuits obtained in Julia to OpenQASM circuits using pytket 2.18.5~\cite{Sivarajah2020} and Qiskit 2.3.1~\cite{qiskit2024}.

The classical compute node has eight Intel Xeon Platinum 8260L sockets at 2.4 GHz, 384 hardware threads, and 6 TB of installed RAM, running Red Hat Enterprise Linux 9.7. 
Quantinuum Helios~\cite{Ransford2026} is a trapped-ion quantum processor comprising 98 physical qubits encoded in the hyperfine states of \(^{137}\mathrm{Ba}^{+}\) ions.
Its quantum charge-coupled device (QCCD) architecture transports ions between eight interaction zones, enabling all-to-all connectivity, with four zones supporting two-qubit gates. 
The native gate set comprises single-qubit rotations and \(ZZ\) interactions, including parametrized-angle gates. 
Helios supports mid-circuit measurement, qubit reuse, and real-time
classical control and conditional branching, supported by the quantum programming language \texttt{guppy}~\cite{Koch2024,Koch2025}. 
Typical single- and two-qubit gate infidelities are \(3\times10^{-5}\) and \(8\times10^{-4}\), respectively, averaged over operational zones, and a typical state-preparation-and-measurement error of \(5\times10^{-4}\)~\cite{quantinuum_helios_datasheet}.
Our hardware measurements are performed with the ternary scheme to herald leakage.

\subsection{Linear solve, integration, and parallel evaluation}

At each imaginary-time step, we evaluate the approximate Gram matrix
$\widetilde M\in\mathbb R^{p\times p}$ and evolution vector
$\widetilde{\vv}\in\mathbb R^p$ for the $p$ circuit angles $\thv$.
The tildes denote quantities evaluated with truncated Pauli propagation.
All main-text results use coefficient truncation without a Pauli-weight cutoff.

To reduce sensitivity to nearly dependent circuit derivative directions, we use Tikhonov-regularized least squares~\cite{Zoufal2021}. At each imaginary-time step, we minimize
$\|\widetilde M\boldsymbol x-\widetilde{\vv}\|_2^2
+\lambda\|\boldsymbol x\|_2^2$
over real parameter velocities $\boldsymbol x$, giving
\begin{equation}
 \left(\widetilde M^{\mathsf T}\widetilde M+\lambda\id\right)
 \dot{\thv}
 =\widetilde M^{\mathsf T}\widetilde{\vv}.
 \label{eq:appregularizedupdate}
\end{equation}
The penalty acts on the parameter velocities, and we keep
$\lambda=10^{-6}$ fixed throughout the reported evolutions.

Each trajectory begins at $\thv_0=\boldsymbol 0$, corresponding to the Bell-pair purification at inverse temperature $\beta_0=0$. We integrate once to the prescribed target temperature using explicit Euler, saving the angles after every step,
\begin{equation}
 \thv_{k+1}
 =\left(\thv_k+\Delta\tau\,\dot{\thv}_k\right)\bmod 2\pi,
 \qquad
 \beta_{k+1}=\beta_k+2\Delta\tau.
 \label{eq:appeulerupdate}
\end{equation}
Here $\dot{\thv}_k$ is evaluated at the current angles, and the modulo operation acts componentwise. Shifting one Pauli-rotation angle by $2\pi$ changes the circuit only by a global sign, so wrapping preserves the physical state. The default step used for all main-text results is $\Delta\tau=0.05$, corresponding to $\Delta\beta=0.1$ because $\beta=2\tau$. Reaching a target $\beta$ requires $\beta/(2\Delta\tau)$ steps from $\beta_0=0$, giving 50 steps for $\beta=5$ at the default step size. The step-size studies below also use $\Delta\tau=0.1,0.02,0.01$.

Generator propagations and weighted Hamiltonian-term propagations are independent tasks distributed over single-threaded worker processes. We use 128 workers for the one-dimensional studies, 96 for the $7\times7$ lattice, and 16 for the $4\times4$ validation, with 32 master threads in each case. Each Hamiltonian term is truncated separately, so distributing the terms across workers does not change their individual approximations. The fixed-worker timing comparisons apply to the reported scalability runs, with compilation and observable-evaluation jobs treated separately.

Before timing each trajectory, we perform one untimed warm-up step at $\thv=\boldsymbol 0$ and discard its output so that Julia compilation is excluded from the measured step times. Reported evolution times therefore cover only the timed propagation, Gram assembly, and parameter-update steps, excluding one-time session initialization, checkpoint writing, and subsequent observable evaluation. For the 128-worker allocation used in the one-dimensional studies, as measured in the 50-site Ising benchmark, this initialization totals approximately 4 minutes, consisting of about 14 seconds for Julia and worker startup, about 230 seconds for loading packages across workers from the shared file system, and about 9 seconds for the discarded warm-up step. During each timed step, we record wall times for generator propagation, Hamiltonian propagation, and Gram assembly, together with retained-term counts and memory diagnostics for the master and worker processes.

We evaluate observables after evolution from the saved circuit angles. For energy, each weighted Hamiltonian term is propagated separately through the circuit at a separate measurement threshold $c_{\mathrm{meas}}$.

\subsection{Exact and approximate references for validation}\label{app:validation}

We choose benchmarks with accessible classical references so that the prepared energies and correlations can be assessed independently of PP-VarQITE. We use exact solutions where available and independent thermal matrix product state (MPS) calculations otherwise. The MPS construction and convergence checks are given in Appendix~\ref{app:mps}. All Hamiltonians use Pauli operators and open boundary conditions, with $n$ denoting the number of physical sites.

\subsubsection{One-dimensional transverse-field Ising model}

For the Ising chain, we use
\begin{equation}
 H_{\mathrm{1D}}
 =J\sum_{i=1}^{n-1}Z_iZ_{i+1}
 +h_x\sum_{i=1}^{n}X_i,
 \qquad J=-1,\quad h_x=-0.5.
 \label{eq:appreference1d}
\end{equation}
The Jordan--Wigner transformation maps this Hamiltonian to a quadratic fermionic model~\cite{Pfeuty1970}, providing exact references for the thermal energy and longitudinal two-point correlations at every inverse temperature $\beta$, as well as the ground-state energy. We compare both energy and correlations
$C_{ij}=\langle Z_iZ_j\rangle$.
Correlations between non-neighboring sites test observables absent from the Hamiltonian used to compute the imaginary-time evolution update, allowing us to assess the correlation structure beyond agreement in energy.

\subsubsection{Frustrated \texorpdfstring{$J_1$--$J_2$}{J1-J2} chain}

The frustrated chain has Hamiltonian
\begin{equation}
 H_{J_1J_2}
 =J_1\sum_{i=1}^{n-1}\boldsymbol\sigma_i\cdot\boldsymbol\sigma_{i+1}
 +J_2\sum_{i=1}^{n-2}\boldsymbol\sigma_i\cdot\boldsymbol\sigma_{i+2},
 \label{eq:appreferencej1j2}
\end{equation}
where
$\boldsymbol\sigma_i\cdot\boldsymbol\sigma_j=X_iX_j+Y_iY_j+Z_iZ_j$.
We set $J_1=1$ and $J_2=0.5$. For even $n$, the product of singlets on bonds $(1,2),(3,4),\ldots$ is an exact ground state with energy
$E_0=-3J_1n/2$~\cite{Majumdar1969}.

We use thermal MPS references to assess energy and dimerisation along the imaginary-time trajectory. Writing
$b_i=\langle\boldsymbol\sigma_i\cdot\boldsymbol\sigma_{i+1}\rangle$,
the dimerisation is
$D=\overline b_{\mathrm{even}}-\overline b_{\mathrm{odd}}$,
where the bars denote averages over even and odd nearest-neighbor bonds, respectively. The singlet product has $D=3$, providing a structural check on the approach to the exact ground-state limit.

\subsubsection{Two-dimensional transverse-field Ising model}

For the square lattice, we use
\begin{equation}
 H_{\mathrm{2D}}
 =J\sum_{\langle i,j\rangle}Z_iZ_j
 +h_x\sum_iX_i,
 \qquad J=-1,\quad h_x=-0.5,
 \label{eq:appreference2d}
\end{equation}
where $\langle i,j\rangle$ denotes nearest-neighbor sites. We compare energy and longitudinal two-point correlations with independent thermal MPS references. As an additional low-temperature energy reference, we use second-order perturbation theory about the zero-field ferromagnetic limit.

At $h_x=0$, a fully polarized ferromagnetic state has energy $-|J|N_b$, where $N_b$ is the number of lattice bonds. Flipping the spin at site $i$ breaks its $z_i$ incident bonds and costs $2|J|z_i$. The transverse field couples the ferromagnetic state to these single-spin-flip states, giving the second-order estimate
\begin{equation}
 E_0^{(2)}
 =-|J|N_b-\sum_i\frac{h_x^2}{2|J|z_i}.
 \label{eq:app2dperturbation}
\end{equation}
For $7\times7$, $N_b=84$, with four sites of coordination $z_i=2$, twenty with $z_i=3$, and twenty-five with $z_i=4$. Equation~\eqref{eq:app2dperturbation} then gives
$E_0^{(2)}/49\simeq-1.75234$.
On $3\times3$, the same expression gives $-1.38310$ per site, compared with the exact value $-1.38324$. This small-system agreement checks the perturbative estimate but does not establish a rigorous uncertainty for $7\times7$.

\clearpage
\section{Tensor network references and circuit compilation}\label{app:mps}

We use purified matrix product states (MPS) both as independent thermal references and as inputs to circuit compilation. The calculations concern the frustrated chain with $J_1=1$, $J_2=0.5$, and the square-lattice transverse-field Ising model with $J=-1$, $h_x=-0.5$, using Pauli operators and open boundary conditions. The circuit comparison uses 50 physical sites for the chain and a $7\times7$ lattice, corresponding to 100 and 98 purification qubits, respectively.

\subsection{Thermal MPS construction}\label{app:mpsthermal}

Each MPS site groups a physical qubit $d_i$ and its ancilla $a_i$, giving local dimension four. For $n$ physical sites, the target purification is~\cite{Feiguin2005}
\begin{equation}
 \ket{\Psi_\beta}
 =\sqrt{\frac{2^n}{Z_\beta}}
 \left(e^{-\beta H/2}\otimes\id_A\right)\ket\Phi,
 \qquad
 \ket\Phi=\bigotimes_{i=1}^n
 \frac{\ket{00}_{d_i a_i}+\ket{11}_{d_i a_i}}{\sqrt2},
 \label{eq:apptnmpspurification}
\end{equation}
where $H$ acts on the physical register and $Z_\beta=\Tr(e^{-\beta H})$. At $\beta=0$, the pair-site MPS is a product state with bond dimension one.

We evolve this state using second-order Trotter time-evolving block decimation~\cite{Vidal2004}, with $\Delta\tau=0.05$ and $\beta=2\tau$. For the chain, each step applies half-step gates over the nearest-neighbor bonds and then the next-nearest-neighbor bonds, followed by the reversed sequence. For the lattice, transverse-field half steps surround one full pass of the commuting $ZZ$ gates. We reverse the order of that pass on alternating steps.

The chain follows physical-site order, while the lattice follows row-major order, making vertical interactions range seven in the pair-site MPS. For nonadjacent interactions, swaps bring the sites together before applying the gate and restore their order afterwards. Each two-site gate and swap is followed by an SVD truncation in canonical gauge. We use a relative discarded-weight tolerance of $10^{-9}$, defined through the ratio of discarded to total squared singular values, together with a hard bond-dimension cap. The cap can require discarding additional weight. We use real arithmetic and renormalize after every time step.

\begin{figure}[htbp]
 \centering
 \includegraphics[width=\textwidth]{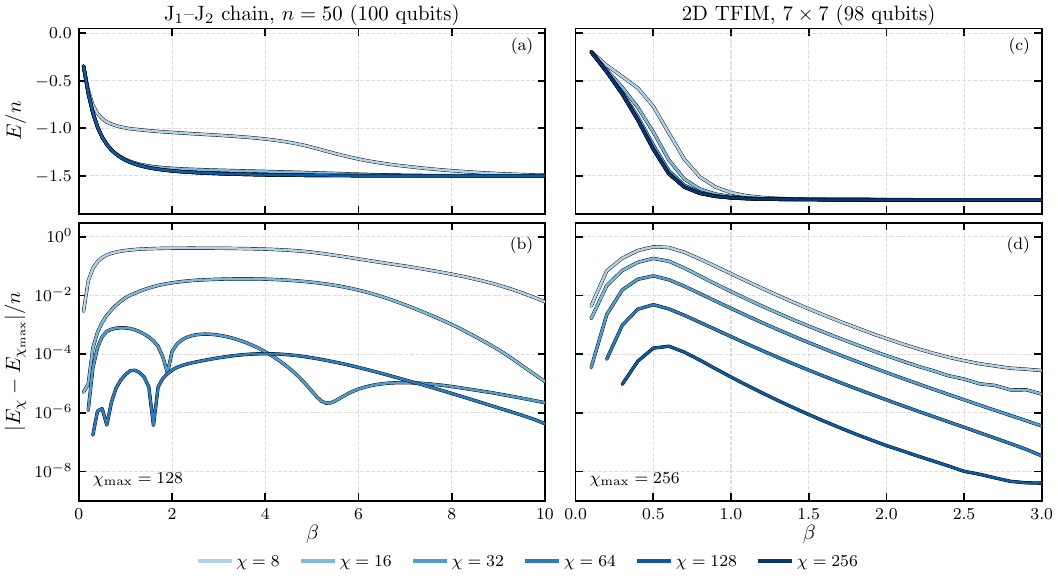}
 \caption{\textbf{Bond-dimension convergence of the thermal MPS references.}
 Panels (a,b) show the 50-site $J_1$--$J_2$ chain and panels (c,d) the $7\times7$ transverse-field Ising model.
 The upper panels give energy per physical site, and the lower panels give
 $|E_\chi-E_{\chi_{\max}}|/n$,
 with $\chi_{\max}=128$ for the chain and 256 for the lattice, the largest bond dimensions covering each full temperature interval.
 Darker blue denotes larger $\chi$, with the same colors used in both models.
 Each curve is independently evolved at its indicated bond-dimension cap, using $\Delta\tau=0.05$ and relative discarded-weight tolerance $10^{-9}$.
 Sharp dips in the lower panels arise from sign changes of the energy difference.}
 \label{fig:mpsconvergence}
\end{figure}

The reference trajectories use $\chi_{\mathrm{ref}}=128$ for the chains and 256 for the lattice, extending to $\beta=10$ and 3, respectively. Energy is obtained by contracting the MPS with the Hamiltonian matrix product operator, while bond and longitudinal pair correlations are evaluated by canonical sweeps. At fixed $\Delta\tau=0.05$, increasing $\chi$ from 64 to 128 for the 50-site chain and from 128 to 256 for the lattice changes the energy by less than $2\times10^{-4}$ per site over the plotted temperature ranges. The bond-dimension sensitivity persists over a broader temperature interval in the frustrated chain, whereas it is concentrated near intermediate temperatures and decreases on further imaginary-time evolution in the lattice, as shown in Fig.~\ref{fig:mpsconvergence}.

\subsection{Compilation into state preparation circuits}\label{app:mpscompile}

At each selected temperature, we compress the reference MPS to a source bond dimension $\chi$ and renormalize it. The compilation inputs are therefore compressed reference states, rather than states evolved throughout imaginary time at the smaller bond dimension. We distinguish compression error from the additional error introduced by approximate circuit compilation.

Before compilation, an SVD with cutoff $10^{-12}$ splits each pair tensor into physical and ancillary qubits. The resulting $N=2n$ qubit MPS has order $d_1,a_1,d_2,a_2,\ldots$. Its bonds are bounded by $\chi$ between pairs and $2\chi$ within a pair. We convert these states to circuits using \texttt{mps-to-circuit}~\cite{MPStoCircuit2025}.

\subsubsection{Exact compilation}

In the exact construction, we left-canonicalize the qubit MPS, pad virtual dimensions to powers of two, and complete the tensor isometries to unitaries~\cite{Lin2021}. Applying these unitaries sequentially from the last qubit to the first prepares the supplied MPS. For $\chi=2^a$, the bulk gates alternate between $(a+1)$- and $(a+2)$-qubit unitaries, explaining the growing synthesis cost with source bond dimension. Exact compilation preserves the selected source state in the ideal construction, while retaining its compression and thermal-reference errors. 

\subsubsection{Approximate compilation}

We use successive disentangling to construct $K$ layers of two-qubit gates~\cite{Ran2020}. At stage $k$, we compress a copy of the current residual state to qubit bond dimension two and compile that copy exactly into a staircase $U_k$. We then apply $U_k^\dagger$ to the residual state and repeat. The resulting state preparation circuit is $ \ket{\Psi_K} =U_0U_1\cdots U_{K-1}\ket{0}^{\otimes N}$. Each staircase contains $N-1$ two-qubit gates and one single-qubit gate. We use this construction without subsequent variational gate optimization. Residual-state updates use a relative discarded-weight tolerance of $10^{-3}$. The unitary completions use random orthogonal columns. These columns leave the exact source state unchanged but can affect the residual updates after the first approximate layer. We fix a random seed for each compilation job and assess sensitivity through separate five-seed checks.

\subsection{Circuit evaluation and resource conventions}\label{app:resource-conventions}

We evaluate the untranspiled circuits using the Qiskit Aer MPS simulator~\cite{qiskit2024} with bond limit 1024, which remains inactive for the reported circuits. Energies and correlations are obtained by contracting the resulting MPS. Alongside the signed energy-density deviation
$(E-E_\beta^{\mathrm{ref}})/n$,
we check the difference between mean even- and odd-bond exchange correlations for the chain, and
$\langle(\sum_i Z_i)^2\rangle/n^2$
for the lattice. We also assess the largest individual bond-correlation or longitudinal pair-correlation discrepancy.

For resource counting, Qiskit~\cite{qiskit2024} decomposes the MPS circuits into single-qubit gates and CNOTs at optimization level zero. Each CNOT counts as one $R_{ZZ}(\pi/2)$ interaction, and depth counts CNOT layers under all-to-all connectivity. For generic two-qubit gates, the approximate construction gives
$3K(N-1)$ two-qubit gates and $3(N-1)+6(K-1)$ two-qubit gate depth. We provide the final gate counts and circuit depth along with the data from the PP-VarQITE approach in Appendix~\ref{app:circuit-resources}.

\clearpage
\section{Extended numerical results}
\label{app:extended-numerics}

We examine how the numerical settings and circuit architecture affect the accuracy and classical cost of PP-VarQITE. For both chain models, we vary the integration step, number of layers, and propagation threshold over system sizes from 10 to 50 physical sites. These comparisons distinguish deviations that respond to numerical refinement from those that persist along the variational trajectory.

\subsection{Impact of imaginary-time step size}

\begin{figure}[htbp]
    \centering
    \includegraphics[width=\linewidth]{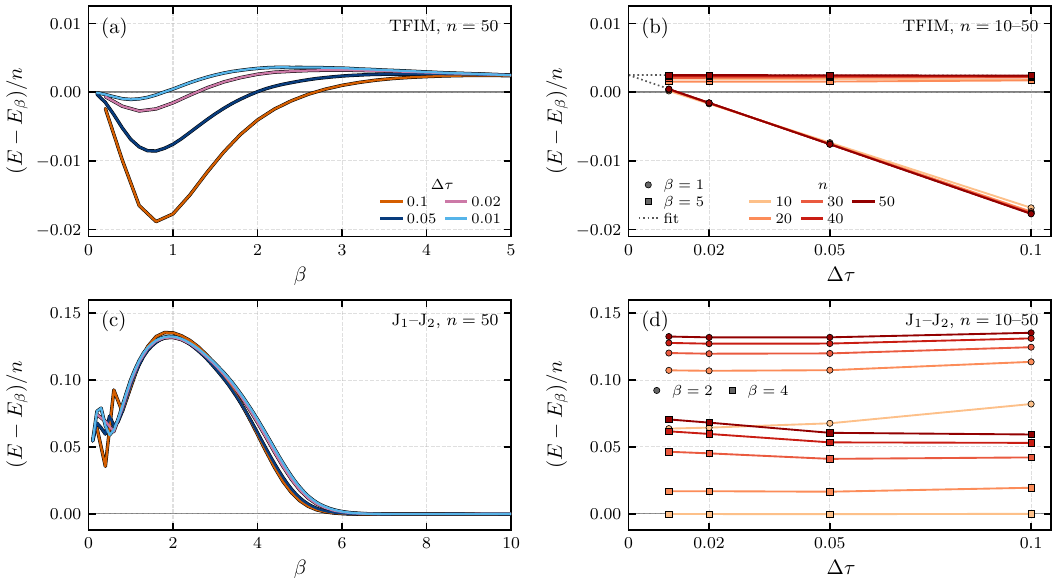}
    \caption{\textbf{Dependence of the thermal energy error on the imaginary-time step size.}
Both chains use OBC, $L=2$, and $c_{\min}=10^{-4}$.
(a,c) Signed energy-density error $(E-E_\beta)/n$ at $n=50$ for the TFIM and frustrated chain, respectively.
Orange, dark blue, pink, and light blue denote $\Delta\tau=0.1,0.05,0.02,0.01$, with $\Delta\beta=2\Delta\tau$.
References are exact for the TFIM and thermal MPS for the frustrated chain.
(b,d) The same errors versus $\Delta\tau$ for $n=10,20,30,40,50$, from light orange to dark red.
Circles and squares denote $\beta=1,5$ in (b) and $\beta=2,4$ in (d).
Dotted lines in (b) are linear fits at $n=50$, extrapolated to zero step size.
The gray horizontal lines mark zero error.}
    \label{fig:dt-ablation}
\end{figure}

We first assess the contribution of the Euler integration step in Fig.~\ref{fig:dt-ablation}. For the Ising chain, the early-time energy deviation depends approximately linearly on the step size, whereas the late-time offset changes little over the tested range. The movement of the zero crossing with the step size also shows that a small absolute energy error can result from cancellation between integration bias and the remaining deviation from the thermal reference. For the frustrated chain, reducing the step size leaves the intermediate-time energy lag largely unchanged and shifts its eventual recovery only modestly. The lag in this model therefore persists beyond the integration error tested here.

\clearpage
\subsection{One-dimensional transverse-field Ising model extended numerical results}

\begin{figure}[htbp]
    \centering
    \includegraphics[width=\linewidth]{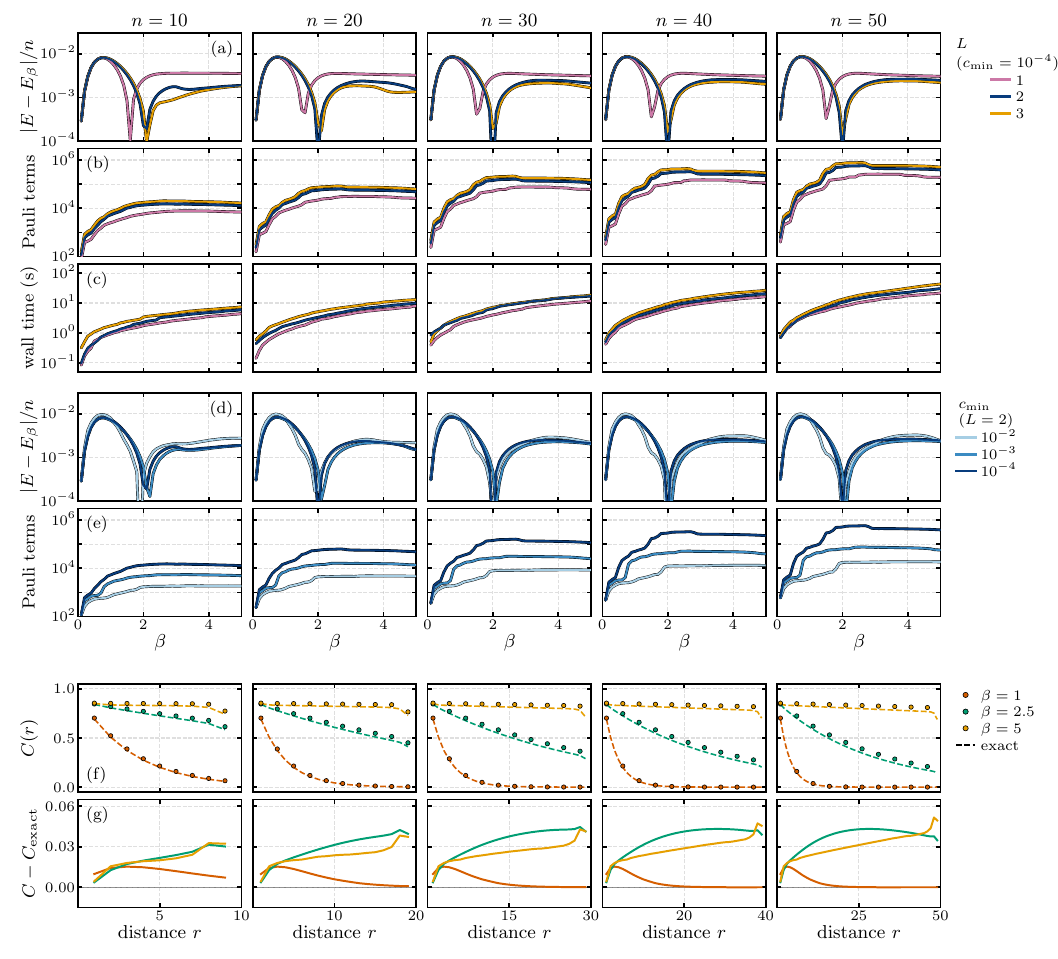}
    \caption{\textbf{Layer and truncation studies for the Ising chain.}
Columns show $n=10,20,30,40,50$ sites with OBC, $J=-1$, $h_x=-0.5$, and $\Delta\tau=0.05$.
(a--c) Absolute energy-density error relative to the exact thermal state, retained Pauli terms per step, and cumulative evolution time for $L=1,2,3$ at $c_{\min}=10^{-4}$, in pink, dark blue, and gold.
(d,e) Energy error and retained terms at $L=2$ for $c_{\min}=10^{-2},10^{-3},10^{-4}$, from light to dark blue.
Times use 128 workers and exclude initialization.
(f) Correlations $C(r)=\langle Z_1Z_{1+r}\rangle$ for $L=2$ and $c_{\min}=10^{-4}$, with orange, green, and gold markers for $\beta=1,2.5,5$ and dashed exact references.
(g) The corresponding signed differences $C-C_{\mathrm{exact}}$.
Dips in the absolute energy error can arise from sign changes.}
    \label{fig:tfim1d-ablation}
\end{figure}

In this subsection, we now examine the layer and truncation dependence of the Ising-chain results in Fig.~\ref{fig:tfim1d-ablation}. We observe similar energy-error profiles across the tested sizes, with modest improvements from additional layers at larger imaginary times. Tightening the propagation threshold increases the number of retained Pauli terms substantially while changing the energy curves comparatively little. The correlation profiles provide a separate check of the prepared states. We reproduce their distance dependence across the chain, although the predominantly positive deviations reveal a residual overestimate of correlations that is less apparent from the energy alone.

\subsection{One-dimensional \texorpdfstring{$J_1$--$J_2$}{J1-J2} model extended numerical results}

\begin{figure}[htbp]
    \centering
    \includegraphics[width=\linewidth]{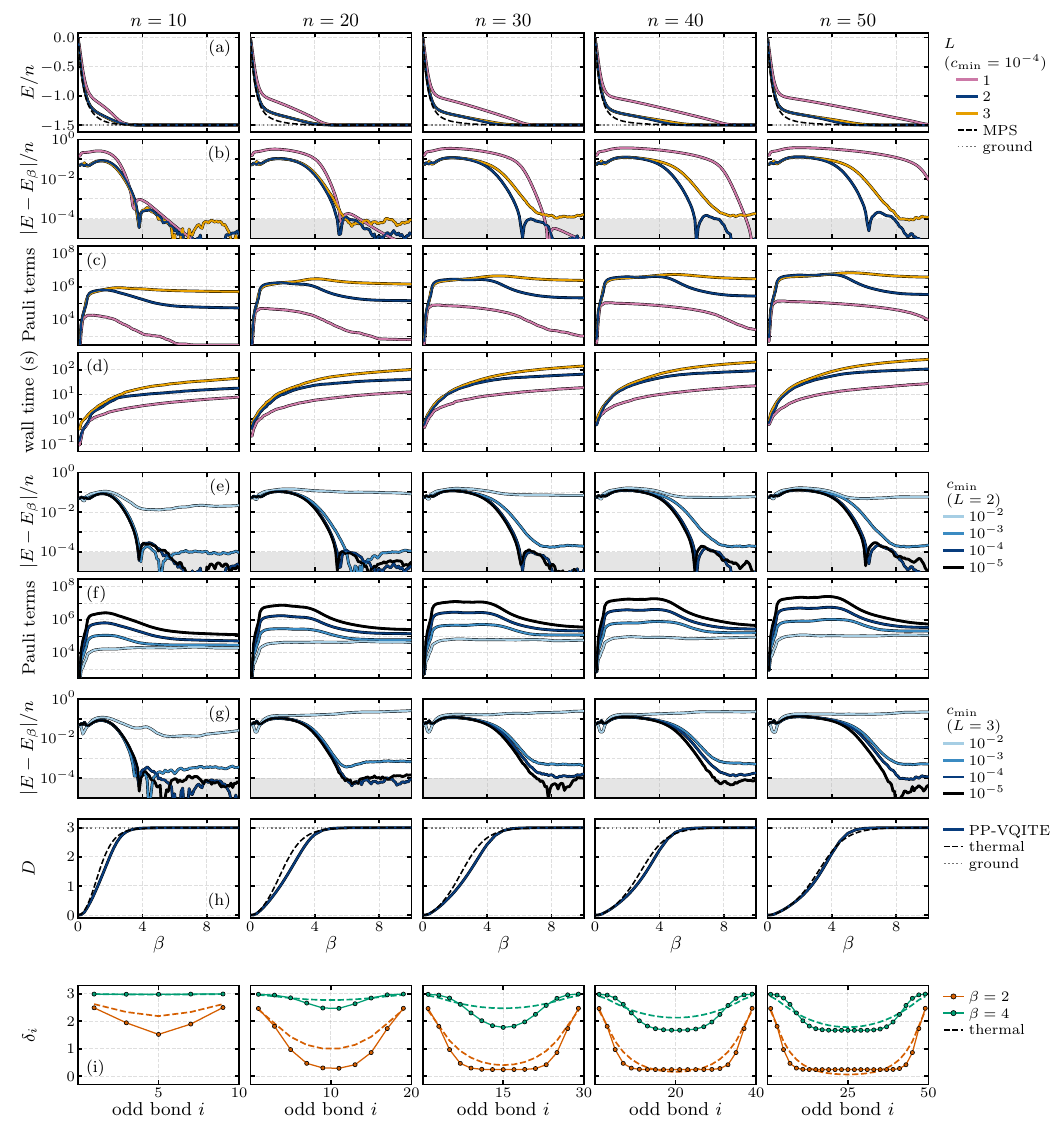}
    \caption{\textbf{Layer and truncation studies for the frustrated chain.}
Columns show $n=10$--50 sites with OBC, $J_1=1$, $J_2=0.5$, and $\Delta\tau=0.05$.
(a--d) Energy density, absolute thermal error, retained Pauli terms, and cumulative evolution time for $L=1,2,3$ at $c_{\min}=10^{-4}$, in pink, dark blue, and gold.
Times use 128 workers and exclude initialization.
(e,f) Error and retained terms at $L=2$, varying $c_{\min}=10^{-2},10^{-3},10^{-4},10^{-5}$ from light blue to black.
(g) The same cutoff study at $L=3$.
Gray bands mark errors below $10^{-4}$ per site.
(h) Chain-averaged dimerisation $D$ for $L=2$, $c_{\min}=10^{-4}$.
(i) Local dimerisation $\delta_i=(b_{i-1}+b_{i+1})/2-b_i$ on interior odd bonds, with $b_i=\langle\boldsymbol\sigma_i\cdot\boldsymbol\sigma_{i+1}\rangle$, at $\beta=2$ (orange) and 4 (green).
Solid curves and markers give circuit predictions, dashed curves give thermal references, and dotted lines give ground-state values.
Energy references use MPS at $\chi=128$, while dimerisation references use exact diagonalisation at $n=10$ and MPS otherwise.}
    \label{fig:j1j2-ablation}
\end{figure}

The frustrated chain is more sensitive to both circuit depth and truncation, as shown in Fig.~\ref{fig:j1j2-ablation}. At larger system sizes, two layers reach the low-temperature energy regime earlier than one layer, while a third layer does not improve the trajectory uniformly. A coarse propagation threshold can prevent convergence to low energies, but tightening the threshold from $10^{-4}$ to $10^{-5}$ does not remove the intermediate-time lag. Within the tested settings, adding layers or retaining more Pauli terms therefore does not guarantee closer agreement with the thermal trajectory.

We also examine the physical content of this lag through the dimerisation profiles in Fig.~\ref{fig:j1j2-ablation}(h,i). The chain-averaged dimerisation approaches the expected singlet pattern with continued evolution, while the local profiles resolve differences in how order develops near the boundaries and in the interior. Agreement in the spatial average can coexist with larger local deviations, motivating the use of spatial observables alongside the energy when assessing finite-temperature state preparation.

\subsection{Scaling in system size}

\begin{figure}[htbp]
    \centering
    \includegraphics[width=\linewidth]{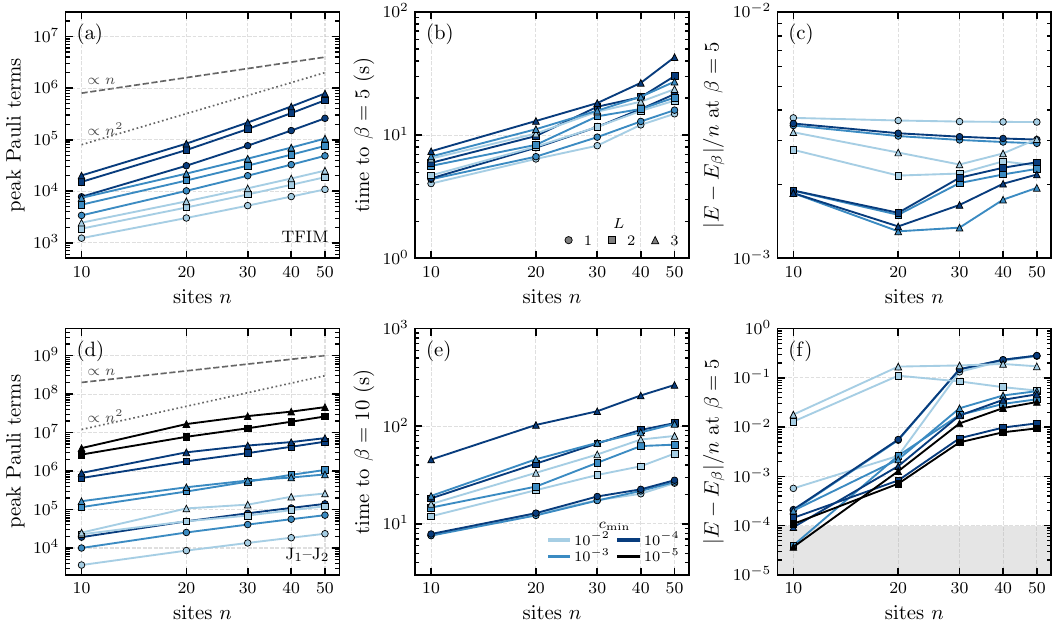}
    \caption{\textbf{Classical cost and energy accuracy versus system size.}
The upper row shows the TFIM and the lower row the frustrated chain, both with OBC and $\Delta\tau=0.05$.
(a,d) Peak retained Pauli terms, with dashed and dotted guides proportional to $n$ and $n^2$.
(b,e) Evolution time to $\beta=5$ for the TFIM and 10 for the frustrated chain, using 128 workers and excluding initialization.
(c,f) Absolute energy-density error at $\beta=5$ relative to the exact TFIM or thermal MPS reference.
Circles, squares, and triangles denote $L=1,2,3$.
Light to dark blue denotes $c_{\min}=10^{-2},10^{-3},10^{-4}$, and black denotes $10^{-5}$ for the frustrated chain at $L=2,3$.
The $10^{-5}$ timings are omitted because those runs shared a node with other jobs.
Gray shading in (f) marks errors below the resolution of $10^{-4}$ per site.}
    \label{fig:scaling-vs-n}
\end{figure}

Over system sizes from $n=10$ to $50$ physical sites, the peak retained Pauli-term count and evolution wall time follow distinct empirical power-law trends for the two models, as shown in Fig.~\ref{fig:scaling-vs-n}(a,b,d,e). In both chains, adding circuit layers or tightening $c_{\min}$ increases the peak term count, which grows approximately as $n^{2}$ for the Ising chain at $c_{\min}=10^{-4}$ and softens to $n^{1.4}$--$n^{1.6}$ at looser cutoffs. Although the frustrated chain retains more terms at tight cutoffs, its peak count grows more slowly with system size, remaining between $n$ and $n^{1.5}$ across all tested layers and cutoffs, including $c_{\min}=10^{-5}$. Because these exponents are fitted over a fivefold range of sizes where local slopes vary, they describe empirical resource growth over the studied range rather than asymptotic scaling laws. Evolution wall times grow approximately linearly with $n$ across all reported settings, reaching at most 43 seconds for the 50-site Ising chain ($\beta=5$) and 265 seconds for the 50-site frustrated chain ($\beta=10$). These wall times exclude the one-time initialization of approximately 4 minutes detailed in Appendix~\ref{app:ablations}, and because every chain length uses the same 128-worker allocation, parallel coordination overhead dominates at smaller sizes.

\subsection{Circuit gate and depth comparison extended results}\label{app:circuit-resources}

We count two-qubit resources in terms of the native $R_{ZZ}(\theta)$ interaction of Quantinuum Helios-1 under all-to-all connectivity. In this convention, each CNOT and each weight-two Pauli rotation costs one two-qubit gate, whereas a weight-three Pauli rotation decomposes into a two-qubit rotation conjugated by two CNOTs and costs three two-qubit gates before further simplification. Table~\ref{tab:resources} summarizes the resulting closed-form scaling expressions and benchmark counts for all three PP-VarQITE ans\"atze alongside the tested MPS compilations, including all fixed Bell-pair and CNOT-network stages. Because each variational sublayer of the PP-VarQITE ans\"atze acts on disjoint qubit pairs, the two-qubit depth per variational layer is independent of system size. The size-dependent contribution to the state preparation depth comes solely from the initial CNOT network, which scales linearly with $n$ for the one-dimensional chains and as $s=\sqrt{n}$ for the breadth-first spanning tree on an $s\times s$ lattice. Complete-circuit depths are smaller than the sum of separately scheduled stage depths because gates acting on disjoint qubits overlap across stage boundaries.

\begin{figure}[!p]
 \centering
 \includegraphics[width=0.92\textwidth]{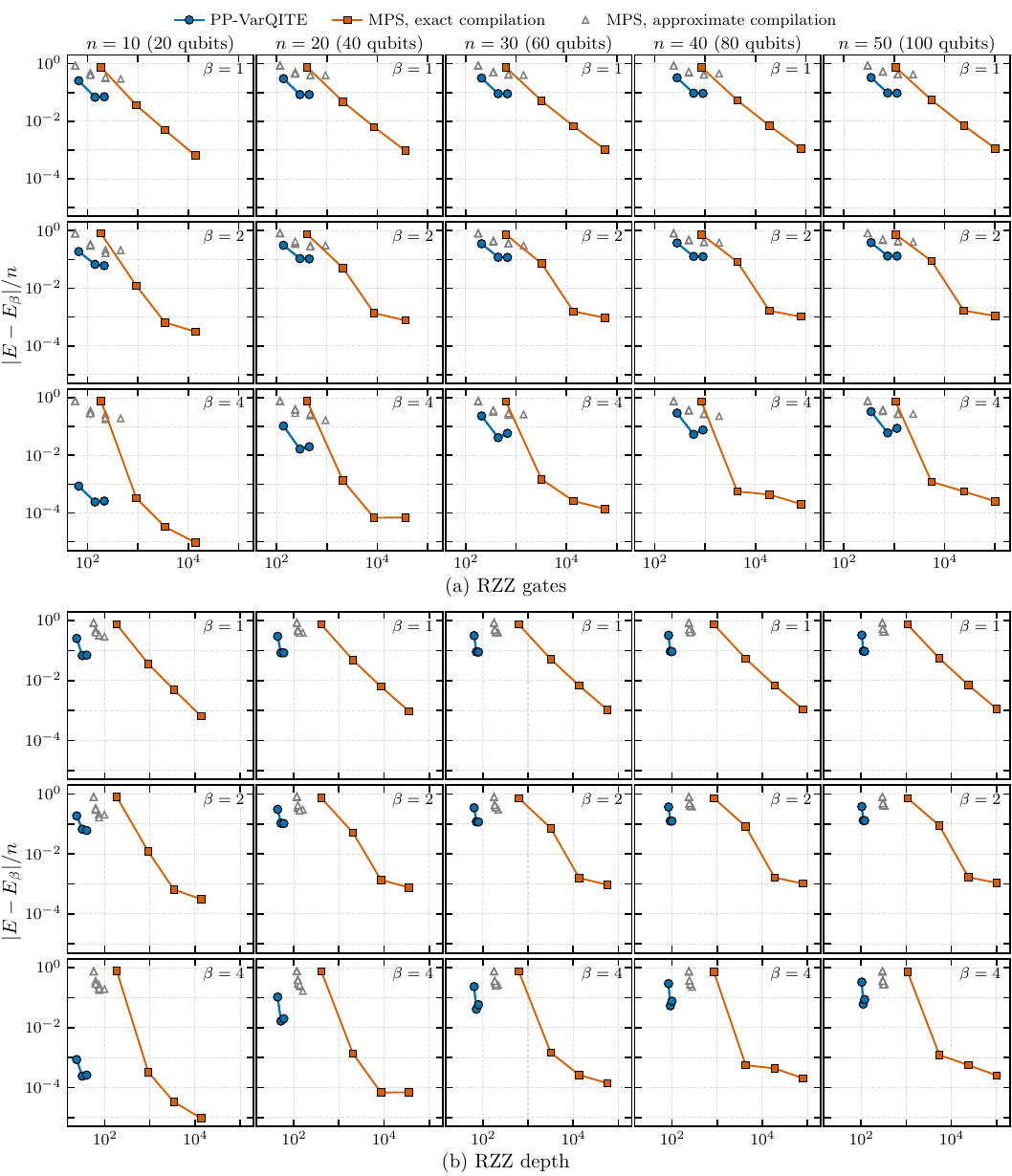}
 \caption{\textbf{System-size scaling of energy accuracy versus circuit gate count and depth for the frustrated chain.}
 Absolute thermal energy-density error $|E-E_\beta|/n$ for the $J_1$--$J_2$ chain ($J_1=1$, $J_2=0.5$) across $n=10,20,30,40,50$ sites ($20$ to $100$ qubits, columns) and inverse temperatures $\beta=1,2,4$ (rows), plotted against (a) native two-qubit $R_{ZZ}$ gate count and (b) native two-qubit $R_{ZZ}$ depth. Blue circles show PP-VarQITE circuits with $L=1,2,3$ layers from left to right. Orange squares show exact compilation of the thermal MPS compressed to bond dimensions $\chi=2,4,8,16$ from left to right. Gray open triangles show approximate MPS compilation with $K=1,2,4$ staircase layers from $\chi=8,16,32$, $K=1,2$ from the uncompressed $\chi=128$ reference, and $K=8$ from $\chi=8$, where markers denote the median over five random unitary completions for $K=2,4$ and vertical bars indicate the full range. All errors are measured relative to the $\chi=128$ thermal MPS reference evolved with $\Delta\tau=0.05$.}
 \label{fig:mps-vs-n}
\end{figure}

We examine how this resource comparison scales with chain length for the frustrated $J_1$--$J_2$ model across $n=10$--$50$ sites and $\beta=1,2,4$ in Fig.~\ref{fig:mps-vs-n}. Across all chain lengths and temperatures in Fig.~\ref{fig:mps-vs-n}(b), the PP-VarQITE circuits remain shallower than every tested MPS compilation. Because both the one-dimensional CNOT ladder and the sequential MPS staircases scale linearly with $n$, their depth ratio approaches a constant factor as the system grows while the absolute number of saved gate layers increases linearly with $n$. In gate count, shown in Fig.~\ref{fig:mps-vs-n}(a), the two-layer PP-VarQITE circuit sits between the two- and four-layer approximate MPS staircases but achieves smaller energy errors across all chain lengths and temperatures, even when the approximate compilation is extended to eight staircase layers. At $\beta=4$, the gradual increase in the PP-VarQITE energy error with $n$ reflects the longer imaginary time needed to close the intermediate-temperature lag in larger chains, while a third variational layer offers no systematic improvement over two layers. Matching the accuracy of the two-layer PP-VarQITE circuit across chain lengths requires exact MPS compilation with at least $\chi=4$ for the energy and $\chi=8$ when also matching the chain-averaged dimerisation, both of which incur an order-of-magnitude or larger overhead in gate count and depth.

\begin{figure}[!t]
 \centering
 \includegraphics[width=\textwidth]{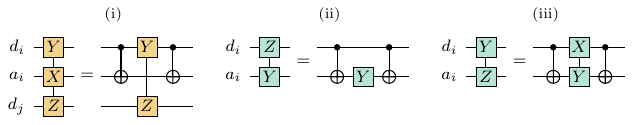}
 \caption{\textbf{Exact changes of Bell frame reduce rotation weight.} Each connected Pauli-letter group denotes a rotation at the same angle on both sides of an identity. (i) Conjugating an orange three-qubit bond rotation by the Bell CNOT gives a two-qubit rotation. (ii) A green site rotation becomes a single-qubit rotation. (iii) The other site rotation remains a two-qubit rotation. Neighboring frame-change CNOTs cancel when successive rotations use the same frame, so the cost reduction depends on the complete sequence.}
 \label{fig:frameidentities}
\end{figure}

For the 2D TFIM ansatz, where the spanning-tree depth grows only as $\sqrt{n}$ while sequential MPS staircases still scale linearly with $n$, we further reduce the as-built gate count and depth before hardware compilation through an exact change of local Bell frame across groups of rotations. Conjugating a Pauli rotation $R_P(\theta)$ by the on-site Bell CNOT $B_i=\mathrm{CNOT}(d_i\to a_i)$ gives $B_i R_P(\theta) B_i = R_{B_i P B_i}(\theta)$, preserving the prepared state at any rotation angle while altering the generator weight. For the three rotation families touching site $i$ in the lattice ansatz, the conjugation identities in Fig.~\ref{fig:frameidentities} give
\begin{equation}
 \begin{aligned}
 B_i(Y_{d_i}X_{a_i}Z_{d_j})B_i&=Y_{d_i}Z_{d_j},\\
 B_i(Z_{d_i}Y_{a_i})B_i&=Y_{a_i},\\
 B_i(Y_{d_i}Z_{a_i})B_i&=X_{d_i}Y_{a_i}.
 \end{aligned}
 \label{eq:appframegenerators}
\end{equation}
In this conjugated frame, the three-qubit bond rotation requires one two-qubit interaction rather than three, and the $Z_{d_i}Y_{a_i}$ site rotation becomes a single-qubit gate. Keeping site $i$ in the conjugated frame across consecutive three-qubit bond and $Z_{d_i}Y_{a_i}$ rotations, deferring the initial Bell-pair CNOT to the first frame entry, and canceling adjacent $B_i$ gates before restoring the standard Bell frame for the remaining bond stages reduces both the two-qubit gate count and the depth of the unpruned $7\times7$ circuits by roughly one quarter, as listed alongside the final Helios-1 compiler output in Table~\ref{tab:resources}. Although all main-text and hardware results retain every variational angle, we also verified that pruning small-angle rotations prior to Bell-frame recompilation can reduce the compiled two-qubit gate count by a further one third to one half at $\beta\ge0.8$ with negligible impact on the energy density.

\begin{table}[htbp]
 \centering
 \footnotesize
 \caption{\textbf{State preparation circuit resources for PP-VarQITE and MPS compilation.} Counts include all initial state preparation stages and retain all variational rotation angles. For the 2D TFIM on an $s\times s$ lattice ($n=s^2$ sites and $N_b=2n-2s$ bonds), the closed-form depth holds for $s\ge4$ ($s+16L$ for odd $s$ and $s+3+16L$ for even $s$), and the benchmark rows distinguish the as-built circuit, exact Bell-frame recompilation, and the \emph{pytket} compilation for Helios-1 at optimization level 2 (at $L=3$, the compiled circuit has $1241$ gates and depth $42$ across the tested temperatures). For MPS compilation on $N=2n$ qubits, $K$ denotes the number of approximate staircase layers and $\chi$ denotes the source bond dimension.}
 \label{tab:resources}
 \setlength{\tabcolsep}{8pt}
 \begin{tabular}{llcccc}
 \hline
 \textbf{Method / Ansatz} & \textbf{Sites ($N_q$)} & \textbf{Setting} & \textbf{Parameters} & \textbf{$R_{ZZ}$ gates} & \textbf{$R_{ZZ}$ depth} \\
 \hline
 \multicolumn{6}{l}{\textit{Closed-form PP-VarQITE scaling (as built)}} \\
 1D TFIM & $n$ ($2n$) & $L$ layers & $(4L+2)n-2L$ & $(4L+3)n-2L-2$ & $n + 4L$ \\
 1D $J_1$--$J_2$ & $n$ ($2n$) & $L\ge2$ & $(8L-4)n-6L+4$ & $(8L-1)n-6L+3$ & $2n + 8L - 4$ \\
 2D TFIM & $s\times s$ ($2n$) & $L$ layers & $2n + L(3N_b+2n)$ & $(2n-1) + L(5N_b+2n)$ & $s + 16L$ \\
 \hline
 \multicolumn{6}{l}{\textit{Benchmark PP-VarQITE circuits}} \\
 1D TFIM & $50$ (100) & $L=1,2,3$ & $298/496/694$ & $346/544/742$ & $54/58/62$ \\
 1D $J_1$--$J_2$ & $50$ (100) & $L=1,2,3$ & $198/592/986$ & $347/741/1135$ & $104/112/120$ \\
 1D $J_1$--$J_2$ (hardware) & $48$ (96) & $L=2$ & $568$ & $711$ & $108$ \\
 2D TFIM, as built & $7\times7$ (98) & $L=2,3,4$ & $798/1148/1498$ & $1133/1651/2169$ & $39/55/71$ \\
 2D TFIM, Bell frame & $7\times7$ (98) & $L=2,3,4$ & $798/1148/1498$ & $845/1243/1641$ & $30/41/52$ \\
 2D TFIM, compiled & $7\times7$ (98) & $L=2,3,4$ & $798/1148/1498$ & $842/1241/1638$ & $30/42/52$ \\
 \hline
 \multicolumn{6}{l}{\textit{Benchmark MPS-compiled circuits}} \\
 Approx.\ MPS ($J_1$--$J_2$) & $50$ (100) & $K=1,2,4,8$ & n.a. & $297/594/1188/2376$ & $297/303/315/339$ \\
 Exact MPS ($J_1$--$J_2$) & $50$ (100) & $\chi=2,4$ & n.a. & $1065/5476$ & $1065/5428$ \\
 Exact MPS ($J_1$--$J_2$) & $50$ (100) & $\chi=8,16$ & n.a. & $24140/102121$ & $23861/100918$ \\
 Approx.\ MPS (2D TFIM) & $7\times7$ (98) & $K=1,2,4$ & n.a. & $291/582/1164$ & $291/297/309$ \\
 Exact MPS (2D TFIM) & $7\times7$ (98) & $\chi=2,4$ & n.a. & $1043/5412$ & $1043/5365$ \\
 Exact MPS (2D TFIM) & $7\times7$ (98) & $\chi=8,16$ & n.a. & $24258/99920$ & $23986/98743$ \\
 \hline
 \end{tabular}
\end{table}

\clearpage
\section{Additional hardware results}
\label{app:hardware-results}

We extend the main-text hardware comparisons with energy estimates
and spatially resolved observables at all measured imaginary times.
For each model, we compare the measurements with predictions for
the executed circuits and with thermal MPS references to assess
hardware deviations alongside the remaining variational discrepancy.
The noiseless circuit predictions are truncated Pauli-propagation
estimates without a hardware-noise model, with numerical precision
assessed through the preceding MPS convergence and PP ablation studies
in Appendices~\ref{app:mps} and~\ref{app:extended-numerics}.

Using the ternary readout scheme on Helios-1 to herald qubit leakage (Appendix~\ref{app:ablations}), we exclude only observable contributions involving leakage-flagged readouts while retaining the remaining unflagged readouts from the same shot, preserving statistical power across 100 attempted shots per basis. For energy estimation, we average the unflagged terms of each Hamiltonian interaction type within each shot and rescale by the total number of terms of that type to form a per-shot contribution in each measurement basis. Each pair or nearest-neighbor bond correlator, including those used to construct the chain-averaged and local dimerisation, is instead averaged over the subset of shots in which both endpoints are unflagged. This per-observable exclusion of leakage-flagged readouts is the only error handling applied to the reported hardware measurements, without dynamical decoupling or additional error mitigation.

We report one sampling standard error for each hardware estimate, retaining covariances between observables extracted from the same shot. For energies, we compute the standard error of the mean from the complete per-shot contributions in each measurement basis and combine the independently sampled bases in quadrature. Because each pair or bond correlator uses its own subset of unflagged shots and multiple terms are measured simultaneously, we estimate uncertainties for dimerisation, correlation maps, and distance-averaged correlations using a delete-one-shot jackknife. These error bars quantify finite-shot sampling uncertainty and exclude systematic effects from device noise or conditioning on unflagged readouts, while neighboring dimerisation values and distinct distance averages remain statistically correlated through shared shots.

For the 48-site $J_1$--$J_2$ chain on 96 qubits, we use a
two-layer state preparation circuit containing 711 two-qubit
gates at a two-qubit depth of 108.
We measure energies that follow
the decrease predicted by the two-layer circuit while remaining
above its noiseless values, as shown in
Table~\ref{tab:hardware-energy-j1j2-48sites}.
At $\beta=2$, the circuit prediction itself lies above the thermal
reference, so both the variational trajectory and hardware execution
contribute to the measured discrepancy. By $\beta=6$, the noiseless
and thermal energies nearly coincide, while a residual hardware
offset remains.

\begin{table}[htbp]
\centering
\caption{\textbf{Energy measurements for the frustrated chain.}
Energy per physical site $E/n$ at the measured inverse temperatures.
Hardware uncertainties denote one sampling standard error,
estimated from 100 shots in each of the $X$, $Y$, and $Z$ bases
with covariances between terms measured together retained.
Noiseless circuit energies are evaluated by truncated Pauli
propagation, while thermal energies are obtained from MPS.
Numerical precision is assessed in the preceding convergence
and ablation studies.}
\label{tab:hardware-energy-j1j2-48sites}
\setlength{\tabcolsep}{12pt}
\begin{tabular}{cccc}
\hline
\multicolumn{4}{c}{
\textbf{48-site $J_1$--$J_2$ Heisenberg chain, 96 qubits}} \\
\hline
Inverse temperature
& Hardware measurement
& Noiseless circuit
& Thermal reference \\
$\beta$
& $E_{\mathrm{hardware}}/n$
& $E_{\mathrm{PP}}/n$
& $E_{\mathrm{MPS}}/n$ \\
\hline
$0.4$ & $-0.9045 \pm 0.0173$ & $-0.9280$ & $-0.9875$ \\
$2.0$ & $-1.2613 \pm 0.0114$ & $-1.3179$ & $-1.4491$ \\
$6.0$ & $-1.4441 \pm 0.0058$ & $-1.4986$ & $-1.4990$ \\
\hline
\end{tabular}
\end{table}

\begin{figure}[htbp]
    \centering
    \includegraphics[width=\linewidth]{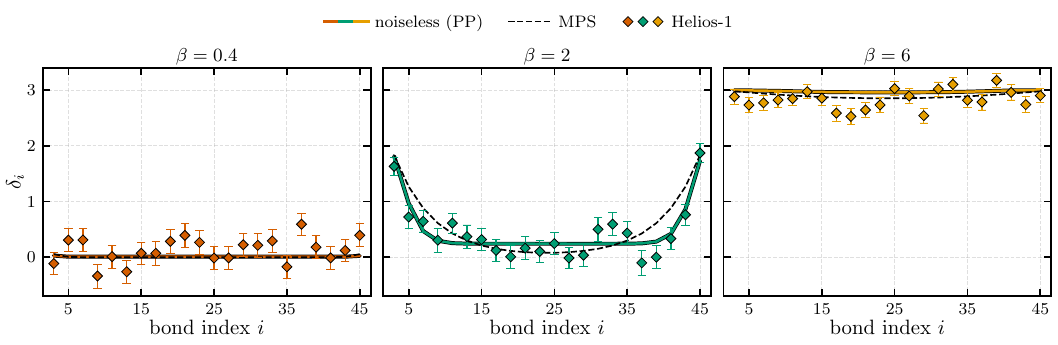}
    \caption{\textbf{Local dimerisation in the 96-qubit hardware state preparation.}
Panels show the 48-site $J_1$--$J_2$ chain with OBC, $J_1=1$, $J_2=0.5$, and $L=2$ at $\beta=0.4,2,6$, from left to right.
Local dimerisation is $\delta_i=(b_{i-1}+b_{i+1})/2-b_i$, where $b_i=\langle\boldsymbol\sigma_i\cdot\boldsymbol\sigma_{i+1}\rangle$, on interior odd bonds $i=3,5,\ldots,45$.
Solid curves give noiseless predictions, black dashed curves give thermal MPS references at $\chi=128$, and diamonds give Helios-1 measurements.
Vermillion, green, and gold identify the three temperatures.
Each run uses 100 attempted shots in each of the $X$, $Y$, and $Z$ bases.
Error bars are one jackknife standard error, with correlated uncertainties across bonds.
Correlator contributions with flagged leaked readouts are excluded, without further mitigation or dynamical decoupling.}
    \label{fig:hardware-j1j2-profiles}
\end{figure}

The local dimerisation profiles in
Fig.~\ref{fig:hardware-j1j2-profiles} reveal how the ordering pattern
develops along this energy trajectory.
At $\beta=0.4$, the expected signal is small compared with the
sampling uncertainty. At $\beta=2$, we resolve stronger dimerisation
near the boundaries and weaker dimerisation in the interior,
following the circuit prediction even where it differs from the
thermal reference. By $\beta=6$, we observe dimerisation throughout
the chain, with its average over the plotted interior bonds retaining
approximately $95\%$ of the noiseless prediction.
The measured ordering pattern therefore approaches the thermal
profile despite the residual energy offset.

For the $7\times7$ TFIM on 98 qubits, we execute three-layer
state preparation circuits containing at most 1241 two-qubit
gates at a two-qubit depth of at most 42 after compilation
for Helios-1.
We observe a similar
intermediate-time lag followed by closer agreement between the
three-layer circuit and the thermal reference, as shown in
Table~\ref{tab:hardware-energy-tfim-7x7}.
The measured energies remain above the noiseless predictions
throughout the sampled trajectory. At $\beta=1.5$, the PP and MPS
energies agree within the numerical precision discussed above,
so the remaining measured offset is associated primarily with
hardware execution.

\begin{table}[htbp]
\centering
\caption{\textbf{Energy measurements for the Ising lattice.}
Energy per physical site $E/n$ at the measured inverse temperatures.
Hardware uncertainties denote one sampling standard error,
estimated from 100 shots in each of the $X$ and $Z$ bases
with covariances between terms measured together retained.
Noiseless circuit energies are evaluated by truncated Pauli
propagation, while thermal energies are obtained from MPS.
Numerical precision is assessed in the preceding convergence
and ablation studies.}
\label{tab:hardware-energy-tfim-7x7}
\setlength{\tabcolsep}{12pt}
\begin{tabular}{cccc}
\hline
\multicolumn{4}{c}{
\textbf{$7\times7$ transverse-field Ising model, 98 qubits}} \\
\hline
Inverse temperature
& Hardware measurement
& Noiseless circuit
& Thermal reference \\
$\beta$
& $E_{\mathrm{hardware}}/n$
& $E_{\mathrm{PP}}/n$
& $E_{\mathrm{MPS}}/n$ \\
\hline
$0.5$ & $-0.8890 \pm 0.0239$ & $-0.9454$ & $-1.2230$ \\
$0.8$ & $-1.4859 \pm 0.0232$ & $-1.5494$ & $-1.6861$ \\
$1.0$ & $-1.6129 \pm 0.0175$ & $-1.7166$ & $-1.7331$ \\
$1.5$ & $-1.6943 \pm 0.0133$ & $-1.7511$ & $-1.7510$ \\
\hline
\end{tabular}
\end{table}

\begin{figure}[htbp]
    \centering
    \includegraphics[width=\linewidth]{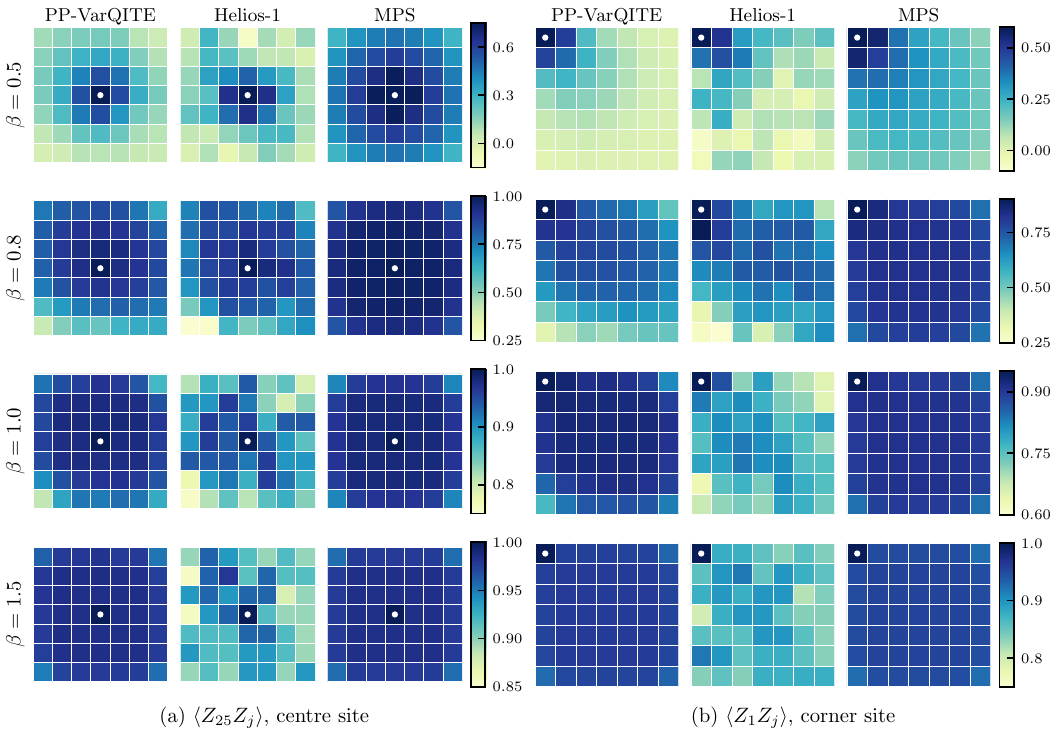}
    \caption{\textbf{Centre- and corner-referenced correlations on the lattice.}
(a) $\langle Z_{25}Z_j\rangle$ from the centre site and (b) $\langle Z_1Z_j\rangle$ from the corner, with white dots marking the reference sites.
Rows show $\beta=0.5,0.8,1.0,1.5$.
Each block compares noiseless PP-VarQITE predictions, Helios-1 measurements, and MPS references at $\chi=256$.
The three maps at each temperature and reference site share a linear colour scale, with darker blue indicating larger correlations.
Scales differ between rows and between (a) and (b).
Hardware estimates use 100 attempted $Z$-basis shots.
Per-site sampling standard errors, omitted from the maps, are approximately $0.08$--$0.10$ at $\beta=0.5$ and $0.02$--$0.08$ at $\beta\geq1$.}
    \label{fig:hardware-tfim2d-mapsr}
\end{figure}

We examine the spatial structure accompanying this energy decrease
through correlations measured from a central site and a corner
in Fig.~\ref{fig:hardware-tfim2d-mapsr}.
At $\beta=0.5$ and $0.8$, the measured maps follow features of
the circuit predictions that differ from the thermal reference,
including the delayed growth of correlations across the lattice.
At larger imaginary times, the circuit predictions approach
the thermal maps while the hardware correlations remain lower.
Comparing the two reference sites reveals spatial variation
in these deviations that is obscured by a single lattice average.

\begin{figure}[htbp]
    \centering
    \includegraphics[width=\linewidth]{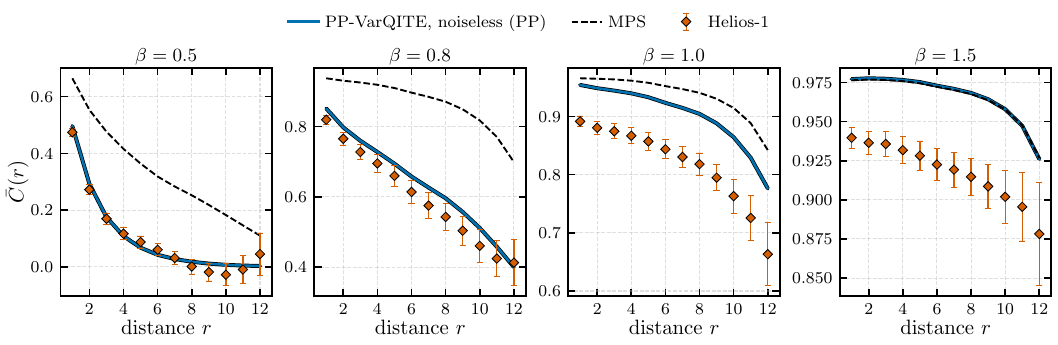}
    \caption{\textbf{Distance-averaged correlations in the 98-qubit hardware state preparation.}
$\bar C(r)$ averages $\langle Z_iZ_j\rangle$ over pairs at Manhattan distance $r$.
Panels show $\beta=0.5,0.8,1,1.5$, each with its own vertical scale. Blue solid curves represent noiseless PP-VarQITE predictions, black dashed curves give MPS references at $\chi=256$, and vermillion diamonds give Helios-1 estimates from 100 attempted $Z$-basis shots.
}
    \label{fig:hardware-tfim2d-cbar}
\end{figure}

To assess how these deviations depend on separation, we average
correlations over pairs at the same Manhattan distance in
Fig.~\ref{fig:hardware-tfim2d-cbar}.
At the two smaller inverse temperatures, the measured curves
follow the noiseless predictions more closely than the thermal
reference. At $\beta=1$, the relative suppression on hardware
increases with distance, whereas at $\beta=1.5$ we retain
approximately $94$--$96\%$ of the noiseless correlations across
the plotted distances. We thus resolve correlations extending
across the lattice after the variational delay has diminished,
with a residual suppression that is approximately independent
of distance at the largest measured inverse temperature.

\end{document}